\documentclass[twocolumn,aps,prl,showpacs,superscriptaddress,groupedaddress]{revtex4}
\usepackage[latin9]{inputenc}
\usepackage{amsmath}
\usepackage{amssymb}
\usepackage{graphicx}
\usepackage{color}
\usepackage{hyperref}
\usepackage[caption=false]{subfig}
\graphicspath{ {images/} }
\usepackage[colorinlistoftodos]{todonotes}
\usepackage{tikz}
\usepackage{circuitikz}
\usepackage{tikz-timing}
\newcommand{\ket}[1]{\left|#1\right\rangle}
\usetikzlibrary{arrows.meta,decorations.pathmorphing,decorations.pathreplacing,positioning,shapes,fadings}

\makeatletter
\@ifundefined{textcolor}{}
{%
 \definecolor{BLACK}{gray}{0}
 \definecolor{WHITE}{gray}{1}
 \definecolor{RED}{rgb}{1,0,0}
 \definecolor{GREEN}{rgb}{0,1,0}
 \definecolor{BLUE}{rgb}{0,0,1}
 \definecolor{CYAN}{cmyk}{1,0,0,0}
 \definecolor{MAGENTA}{cmyk}{0,1,0,0}
 \definecolor{YELLOW}{cmyk}{0,0,1,0}
}

\usepackage{dcolumn}% needed for some tables
\usepackage{bm}% for math
\makeatother

\begin{document}

\title{Universal stabilization of a parametrically coupled qubit}

\author{Yao Lu}
\affiliation{The James Franck Institute and Department of Physics, University of Chicago, Chicago, Illinois 60637, USA}
\author{S. Chakram}
\affiliation{The James Franck Institute and Department of Physics, University of Chicago, Chicago, Illinois 60637, USA}
\author{N. Leung}
\affiliation{The James Franck Institute and Department of Physics, University of Chicago, Chicago, Illinois 60637, USA}
\author{N. Earnest}
\affiliation{The James Franck Institute and Department of Physics, University of Chicago, Chicago, Illinois 60637, USA}
\author{R. K. Naik}
\affiliation{The James Franck Institute and Department of Physics, University of Chicago, Chicago, Illinois 60637, USA}
\author{Ziwen Huang}
\affiliation{Department of Physics and Astronomy, Northwestern University, Evanston, Illinois 60208, USA}
\author{P. Groszkowski}
\affiliation{Department of Physics and Astronomy, Northwestern University, Evanston, Illinois 60208, USA}
\author{Eliot Kapit}
\affiliation{Department of Physics and Engineering Physics, Tulane University, New Orleans, Louisiana 70118, USA}
\author{Jens Koch}
\affiliation{Department of Physics and Astronomy, Northwestern University, Evanston, Illinois 60208, USA}
\author{David I. Schuster}
\affiliation{The James Franck Institute and Department of Physics, University of Chicago, Chicago, Illinois 60637, USA}
\email{David.Schuster@uchicago.edu}

\date{\today }
\begin{abstract}

We autonomously stabilize arbitrary states of a qubit through parametric modulation of the coupling between a fixed frequency qubit and resonator. The coupling modulation is achieved with a  tunable coupler design, in which the qubit and the resonator are connected in parallel to a superconducting quantum interference device.  This allows for quasi-static tuning of the qubit-cavity coupling strength from 12\,MHz to more than 300\,MHz.  Additionally, the coupling can be dynamically modulated, allowing for single photon exchange in 6\,ns.  Qubit coherence times exceeding 20\,$\mu$s are maintained over the majority of the range of tuning, limited primarily by the Purcell effect.  The parametric stabilization technique realized using the tunable coupler involves engineering the qubit bath through a combination of photon non-conserving sideband interactions realized by flux modulation, and direct qubit Rabi driving. We demonstrate that the qubit can be stabilized to arbitrary states on the Bloch sphere with a worst-case fidelity exceeding $80\%$.

\end{abstract}
\maketitle

Dissipation is generally thought of as competing with quantum coherence.  However, under appropriate circumstances dissipation can be engineered and utilized as a resource for coherent quantum control~\cite{Wineland1987LaserCooling,Poyatos1996QuantumIons,Vuletic2000LaserScattering}. Dissipation can be used to generate and stabilize entangled states~\cite{Shankar2013AutonomouslyBits} and many-body phases~\cite{Ma2017AutonomousSolids, Anderson2016EngineeringArrays}. Quantum error correction, one of the main goals in quantum information science, can also be achieved by autonomously stabilizing a manifold of states~\cite{Leghtas2013Hardware-EfficientProtection,Kapit2015PassiveFabric, Kapit2016Hardware-EfficientCircuits} through bath engineering, without the need for active feedback. In superconducting circuit QED, engineered dissipation has been used in conjunction with the Josephson non-linearity of the qubit to achieve stabilization of qubit ~\cite{Leek2009UsingCircuitsb,Murch2012Cavity-AssistedEngineering,Kimchi-Schwartz2016StabilizingQubitsb,Shankar2013AutonomouslyBits} and cavity states~\cite{Holland2015Single-Photon-ResolvedStates}, primitives for autonomous error correction. A more convenient approach to quantum state stabilization, however, may lie in the direct modulation of the coupling between the system and a quantum bath, a task that can be accomplished by using tunable coupler devices ~\cite{Allman2014TunableResonator, Chen2014QubitCoupling, Sirois2015Coherent-stateConversion, McKay2016UniversalBus}.

Tunable coupling elements can mediate interactions while maintaining coherence. They have been used for frequency conversion~\cite{ Zakka-Bajjani2011QuantumStates, Sirois2015Coherent-stateConversion}, quantum logic gates~\cite{Chen2014QubitCoupling, McKay2016UniversalBus}, and are suitable for a variety of tasks in quantum information processing~\cite{Didier2015FastInteraction,Naik2017RandomProcessors} and quantum simulation~\cite{Roushan2016ChiralField}. In this letter, we present a tunable coupling circuit in which a single-junction transmon is coupled to a dissipative bath in the form of a low-Q cavity, via grounding through a shared dc SQUID. We show that the coupling can be tuned over a large dynamic range using magnetic flux, with very little qubit dephasing from flux noise. By parametric modulation of the coupling, we realize both photon conserving red-sideband interactions to transfer single photons~\cite{Beaudoin2012First-orderModulation,Strand2013First-orderQubits}, as well as photon non-conserving blue-sideband interactions~\cite{Wallraff2007SidebandCavity,Leek2009UsingCircuits,Novikov2015RamanSystem} necessary for state stabilization. We present a scheme to parametrically stabilize arbitrary single-qubit states by using the blue-sideband interaction in conjunction with a regular qubit Rabi drive.

\begin{figure}
\subfloat{
\includegraphics{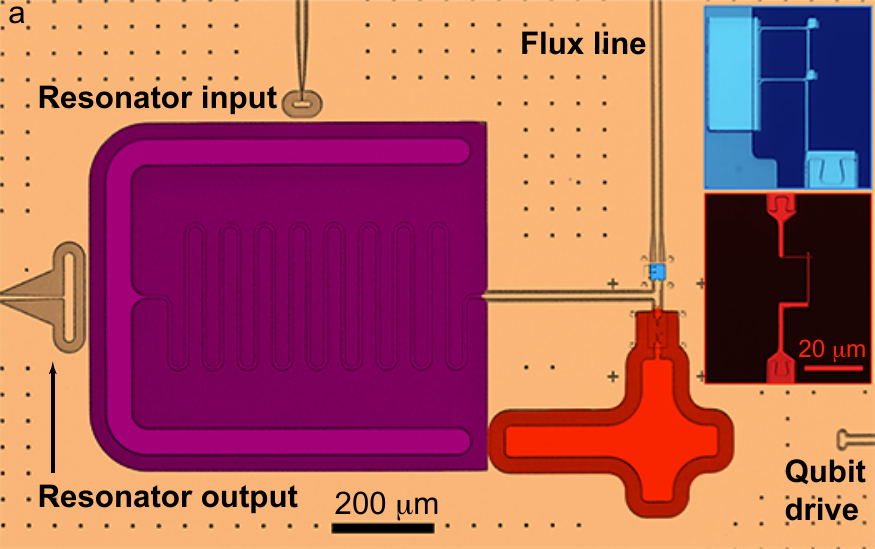}
}\vfill
\subfloat{\includegraphics[scale=0.33]{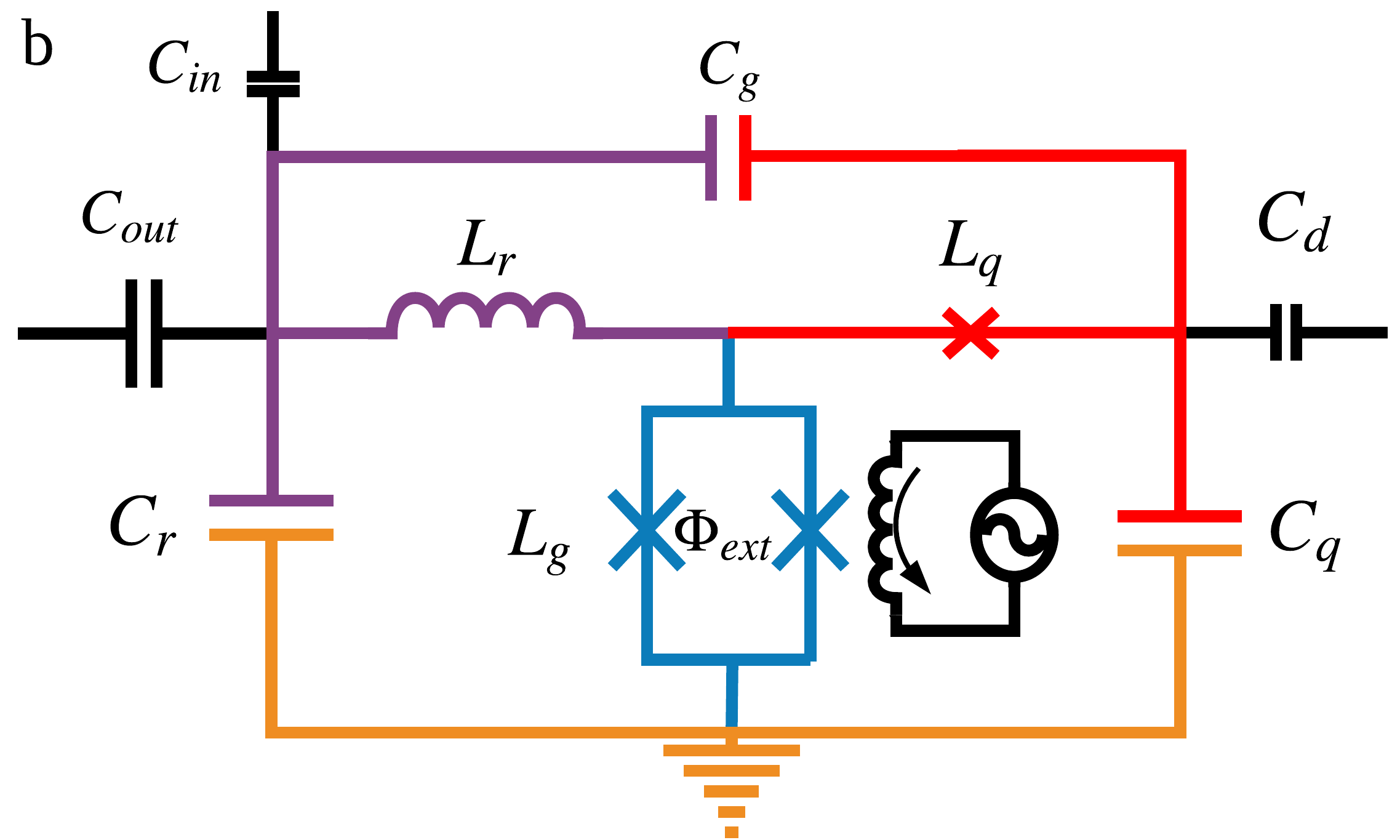}}

\caption{(a) Optical image and (b) circuit diagram of our device. The lumped-element
resonator is formed by a ``C'' shaped capacitor pad and an isolated
meander line inductor. The inductor line protrudes to the common node
where both the qubit Josephson junction and the coupler SQUID loop
are connected. Two voltage ports are placed at the two sides of
the resonator's capacitor pad enabling transmission measurements.
The qubit-cavity coupling strength is tuned with the SQUID-loop flux by modulating the current that flows through the flux line. The qubit can be probed via a separate qubit drive line that is weakly coupled to the qubit's shunting capacitor. Insets show the details of the qubit Josephson junction and dc SQUID loop.
\label{circuit layout}}
\end{figure}

The tunable coupling circuit, shown in Fig.~\ref{circuit layout}, consists of a transmon qubit~\cite{Koch2007Charge-insensitiveBox} and a lumped-element resonator, both grounded at the same node through a dc SQUID. The dc SQUID  acts as a tunable inductor shared between the qubit and the resonator, creating a coupling strength between the two systems proportional to its inductance $L_g=L_{g0}/\left|\cos(\pi\Phi_{\rm ext} /\Phi_0)\right|$, which is controlled by the external flux $\Phi_{\rm ext}$ threading the loop. Previous tunable coupler designs~\cite{Chen2014QubitCoupling,McKay2016UniversalBus} utilized series coupling schemes which are convenient for chains and lattices of qubits or resonators.  By contrast,  the topology of our circuit enables many resonators or qubits to share the same coupler, which is suitable for random access memories ~\cite{Naik2017RandomProcessors}.  
The circuit is described by the effective Hamiltonian,
\begin{align}
\hat{H}= & \omega_{r}\hat{a}^{\dagger}\hat{a}+\frac{\omega_q}{2}\hat{\sigma}_z\nonumber \\
 & -g_{r}(\hat{a}^{\dagger}\hat{\sigma}^{-}+\hat{a}\hat{\sigma}^+)-g_{b}(\hat{a}^{\dagger}\hat{\sigma}^{+}+\hat{a}\hat{\sigma}^-),\label{eq:Hamiltonian}
\end{align}
where
\begin{equation}
g_{r,b}=\frac{L_{g0}}{2\left|\cos(\pi\Phi_{\mathrm{ext}}/\Phi_{0})\right|}\sqrt{\frac{\omega_{r}\omega_{q}}{L_{r}L_{q}}} \mp \frac{C_g}{2}\sqrt{\frac{\omega_{r}\omega_{q}}{C_{r}C_{q}}}\label{eq:grb}
\end{equation}
are the coupling strengths associated with the red and blue sidebands~\cite{Wallraff2007SidebandCavity}. The operators $\hat{a}$ and $\hat{\sigma}^-$ represent the lowering 
operators for the cavity and the qubit mode, and $\omega_{r}$, $\omega_{q}$ are the mode frequencies. The definitions of inductances and capacitances for qubit and resonator can be read off from Fig.~\ref{circuit layout}b. It should be noted that for the Hamiltonian above, the degree of freedom associated with the SQUID coupler has been adiabatically eliminated (see Supplementary Information).  When the coupler is not being driven, the counter-rotating $g_b$ term can usually be dropped from Eq.~\eqref{eq:Hamiltonian}, but by dynamically modulating the inductance via the external flux $\Phi_{\rm ext}$, both red- and blue-sideband interactions can be utilized. Additionally, by balancing the inductive and capacitive terms in Eq.~\eqref{eq:grb}, one can make $g_r$ zero or even negative.

\begin{figure}[t]
\includegraphics[width=3.375in]{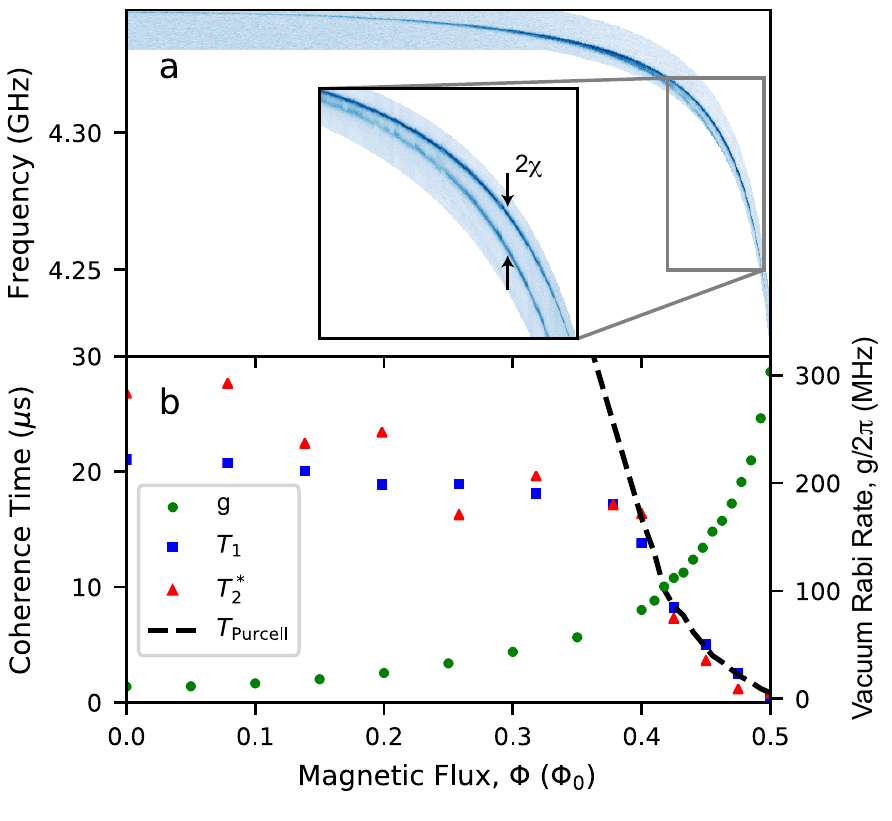}
\caption{(a) Response of qubit as a function of flux through the coupler, showing the insensitivity of the qubit frequency over the entire flux range. (Inset) Number splitting of the qubit peak due to photons in the resonator, used to calibrate the static coupling between the qubit and the resonator. (b) Qubit coherence and qubit-cavity coupling strength as a function of the flux through the coupler. The dephasing time ($T^{*}_{2}$) is comparable to the energy relaxation time ($T_{1}$) over the entire tuning range. The coherence times drop near $\Phi = 0.5\Phi_{0}$ as a result of the Purcell effect due to the strong coupling to the readout resonator, as indicated by the black dashed line. \label{spec_and_coherence}}
\end{figure}

As the qubit itself does not have a SQUID loop, its frequency is only indirectly affected by the modulation of the coupler. We choose $L_{g0}\ll L_{r,\,}L_{q}$ for our device to ensure that the tuning of the qubit and resonator frequencies from the change in the coupler inductance is small. This can be seen in Fig.~\ref{spec_and_coherence}a, where the qubit frequency varies by less than 15 MHz over 80\% of the tuning range, making the qubit nearly immune to flux noise.  Spectroscopy of the qubit can also be used to determine the static coupling strength $g_r(\Phi_{\rm ext})$. Photons in the resonator (from the readout), result in photon-number splitting of peaks~\cite{Schuster2007ResolvingCircuit} with separation $2\chi=g_r^2 \alpha/\Delta(\Delta+\alpha)$ \cite{Koch2007Charge-insensitiveBox}, where $\alpha = -188$\,MHz is the qubit anharmonicity and $\Delta$ is the qubit-cavity detuning. At flux values where the splitting is too small to be resolved, we calibrate $g_r$ by measuring the qubit Rabi rate through the cavity at fixed power (See Supplementary Information for details). We display the tuning of the coupling and the simultaneous robustness of the qubit coherence in Fig.~\ref{spec_and_coherence}b. Both the energy relaxation time $T_1$ and the dephasing time $T_2^{*}$ remain above 20\,$\mu s$ over most of the flux period ($\left|\Phi_{\rm ext}\right|<0.4\Phi_{0}$). Only when the flux approaches half a flux quantum do coherence times start to drop significantly. There the Purcell effect from coupling to the readout resonator, as well as an increased frequency-flux sensitivity, limit the coherence.

\begin{figure}[t]
\begin{tikzpicture}[font={\fontfamily{phv}\selectfont}]
          \node[anchor=north west,inner sep=0] (image) at (0,0)	{\includegraphics[scale=0.96]{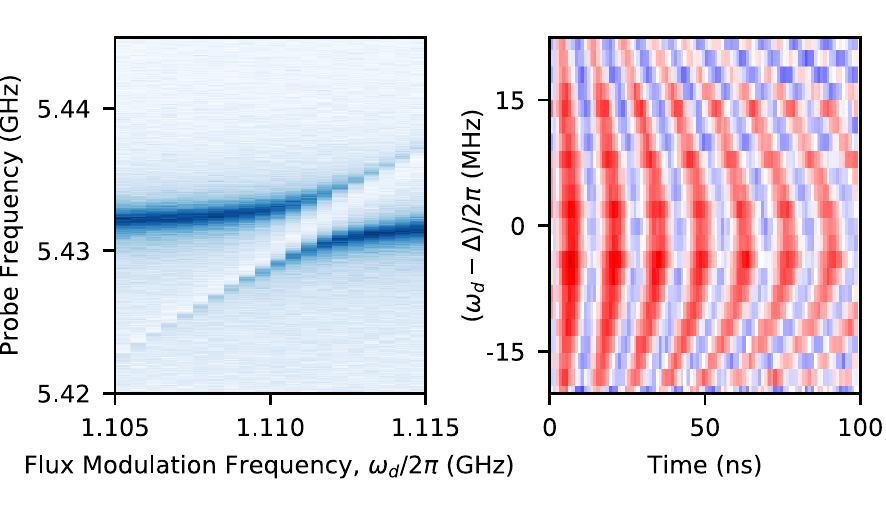}};
          \node[scale=0.8] at (0.7,-0.5) {\textbf{a}};  
          \node[scale=0.8] at (5,-0.5) {\textbf{b}};  
\end{tikzpicture}
\caption{red-sideband interactions probed by applying an rf flux tone to the tunable coupler to generate sidebands. (a) Transmission of the readout resonator as a function of sideband and resonator probe frequency, showing the stimulated vacuum Rabi spitting.  (b) Time domain stimulated vacuum Rabi oscillations between the qubit and resonator, measured as an oscillation of the qubit excited state population, with Rabi frequency of 80 MHz. A single photon is loaded into the qubit with a $\pi$-pulse at the beginning of the sequence. \label{red_sideband}}
\end{figure}

The usefulness of parametric coupling becomes most evident when the qubit-cavity coupling strength is modulated at the qubit-cavity difference or sum frequency. Harmonic modulation of $\Phi_{\rm ext}$ in Eq.~\eqref{eq:grb} turns $g_{r,b}$ into periodic functions with leading order Fourier series expansion as $g_{r,b}(t)=g_{r,b}^{(0)}+g_{r,b}^{(1)}\cos\omega_d t$. Substituting this into Eq.~\eqref{eq:Hamiltonian}, we obtain the red- and blue-sideband Hamiltonians in rotating frames as
\begin{equation}
\hat{H}^{r,b}_{rot}=(\omega'_{r}\mp\omega'_{q}-\chi'\hat{\sigma}_z)\hat{a}^{\dagger}\hat{a}\pm\frac{\omega_d}{2}\hat{\sigma}_z-g_{r,b}'(\hat{a}^{\dagger}\hat{\sigma}^{\mp}+\hat{a}\hat{\sigma}^{\pm}),\label{eq:H_rb_rot}
\end{equation}
valid for flux modulation frequencies, $\omega_d \approx \omega_{r}'\pm (\omega_{q}'+\chi')$, respectively, with fast-oscillating terms abandoned. Here, the primes stand for the dressed basis after diagonalizing the static component of the driven Hamiltonian. At $\omega_d = \omega_{r}'-\omega_{q}'+\chi'$, 
energy pumped into the circuit through the parametric flux drive bridges the gap between the first excited state of the qubit $\left|e0\right\rangle $ and the single-photon Fock state of the cavity $\left|g1\right\rangle $, causing a  splitting of $2g_r'$ due to the red-sideband coupling between the two levels. This is seen as an avoided crossing in the cavity transmission spectrum when the modulation frequency matches the detuning, see Fig.~\ref{red_sideband}a. In the time domain, the red-sideband coupling mediates stimulated vacuum Rabi oscillations which coherently swap a single photon between qubit and resonator~\cite{Beaudoin2012First-orderModulation,Strand2013First-orderQubits}. The oscillation rate, $2g_r'/2\pi\approx 80$ MHz, can be directly seen from Fig.~\ref{red_sideband}b and determines how fast qubit-photon gates can be performed.

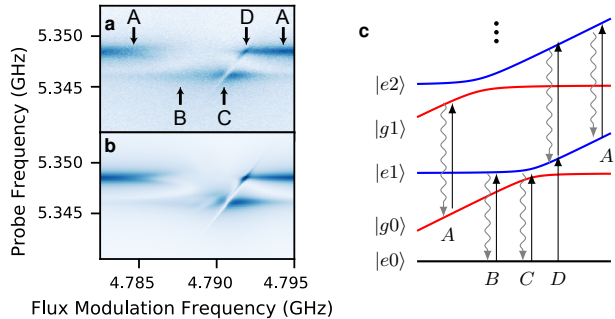
\begin{figure}[t]

\subfloat{
\begin{tikzpicture}[font={\fontfamily{phv}\selectfont}]
          \node[anchor=north west,inner sep=0] (image) at (0,0) [scale=0.9]{\includegraphics[scale=0.95]{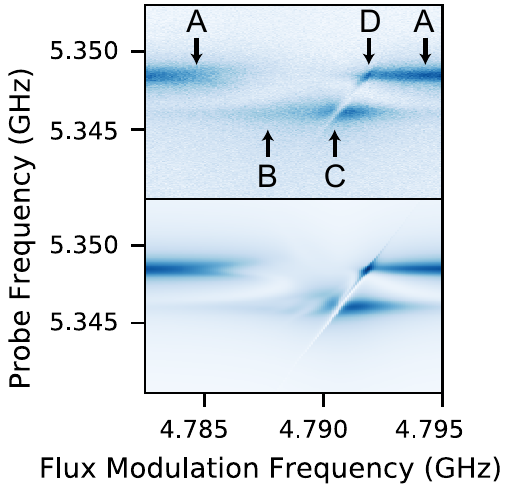}};
          \node[scale=0.8] at (1.4,-0.3) {\textbf{a}};  
          \node[scale=0.8] at (1.4,-1.9) {\textbf{b}};  
%           \node[scale=0.8] at (1.6,-0.3) {\textbf{c}};  
\end{tikzpicture}
}%\hspace*{\fill}
\begingroup
\subfloat{
\begin{tikzpicture}[scale=0.17\textwidth / 1.3cm,font={\fontfamily{phv}\selectfont},level/.style={thin},
          wavy/.style={->,decorate,decoration={snake,amplitude=.35mm,segment length=2mm,post length=1mm},color=gray,>=latex},
          trans/.style={thin,-{Latex[length=1mm,width=1.5mm]},shorten <=2pt,decoration={snake,amplitude=0.4mm}},
          classical/.style={<-, >=latex},
          label/.style={font=\small}]
  \draw[color=black,thick,samples=300,domain=-0.15:0.95,id=avc1] plot (\x, {0.75} );
\draw[color=red,thick,samples=300,domain=-0.15:0.95,id=avc1] plot (\x, {0.75+0.25*((0.5+\x)+1.0-sqrt(4*0.05^2+(0.5+\x-1.0)^2)} );
  \draw[color=blue,thick,samples=300,domain=-0.15:0.95,id=avc2] plot (\x, {0.75+0.25*((0.5+\x)+1.0+sqrt(4*0.05^2+(0.5+\x-1.0)^2)} );
  \draw[color=red,thick,samples=300,domain=-0.15:0.95,id=avc3] plot (\x, {0.75+0.25*((1.8+\x)+2.0-sqrt(4*2*0.05^2+(1.8+\x-2.0)^2)} );
  \draw[color=blue,thick,samples=300,domain=-0.15:0.95,id=avc4] plot (\x, {0.75+0.25*((1.8+\x)+2.0+sqrt(4*2*0.05^2+(1.8+\x-2.0)^2)} );
 \node[scale=1] at (1.5,0.45) {};
   \draw[color=white,thick,samples=300,domain=-0.15:0.95,id=avc1] plot (\x, {0.6} );
\node[scale=0.8] at (-0.3,0.75) {$\left|e0\right\rangle$};
\node[scale=0.8] at (-0.3,1.25) {$\left|e1\right\rangle$};
\node[scale=0.8] at (-0.3,1.75) {$\left|e2\right\rangle$};
\node[scale=0.8] at (-0.3,0.95) {$\left|g0\right\rangle$};
\node[scale=0.8] at (-0.3,1.5) {$\left|g1\right\rangle$};
\draw [wavy] (0.0,1.65) -- (0.0,1.0);
\draw [classical] (0.05,1.65) -- (0.05,1.05);
\node[scale=0.8] at (0.025,0.9) {$A$};
\draw [classical] (0.3,1.25) -- (0.3,0.75);
\draw [wavy] (0.25,1.25) -- (0.25,0.75);
%\draw [wavy] (0.3,1.7) -- (0.3,1.25);
%\draw [classical] (0.2,1.8) -- (0.2,1.1);
\node[scale=0.8] at (0.275,0.65) {$B$};
\draw [classical] (0.5,1.25) -- (0.5,0.75);
\draw [wavy] (0.45,1.25) -- (0.45,0.75);
\node[scale=0.8] at (0.475,0.65) {$C$};
\draw [classical] (0.65,1.35) -- (0.65,0.75);
\draw [classical] (0.65,2.0) -- (0.65,1.35);
\draw [wavy] (0.6,1.95) -- (0.6,1.3);
\node[scale=0.8] at (0.65,0.65) {$D$};
\draw [classical] (0.9,2.1) -- (0.9,1.45);
\draw [wavy] (0.85,2.05) -- (0.85,1.45);
\node[scale=0.8] at (0.925,1.35) {$A$};
\node[scale=0.8] at (-0.45,2.05) {\textbf{c}}; 
% \node[scale=0.8] at (1.5,1.5) {};
\node[draw,circle,fill=black,scale=0.15] at (0.3,2.05) {};
\node[draw,circle,fill=black,scale=0.15] at (0.3,2.1) {};
\node[draw,circle,fill=black,scale=0.15] at (0.3,2.0) {};
  \end{tikzpicture}
}
\endgroup
\caption{Resonator transmission spectroscopy with flux modulation swept across the blue-sideband frequency. Experimental data (a) and  master equation simulations (b) show excellent agreement. (c) Energy level diagram corresponding to Eq.~\eqref{eq:H_rb_rot} provides a map to the spectroscopic features A, B, C and D at different modulation frequencies, indicated by arrows. A: When the flux modulation frequency is far-detuned from the blue-sideband frequency, the qubit stays in its ground state. 
B: The excited state of the qubit is stabilized, causing the cavity to be shifted down by $2\chi$.
C: The crossing of $\left|e1\right\rangle$ and $\left|g0\right\rangle$, manifest as an avoided crossing. The qubit excited state is also maximally stabilized at this frequency due to the resonance of $\left|e1\right\rangle$ and $\left|g0\right\rangle$. D: Enhanced cavity transmission appears as a bright spot, where $\left|e0\right\rangle \rightarrow \left|g0\right\rangle$ and  $\left|g0\right\rangle \rightarrow \left|g1\right\rangle$ transition energies are equal. The asymmetry of the un-shifted cavity peak line centered at the blue-sideband frequency is likely due to interactions between higher levels $\left|g,n\right\rangle \rightarrow \left|e,n+1\right\rangle$.} 
\label{blue sideband plot}
\end{figure}

\begin{figure*}[t] 
\subfloat{\begin{tikzpicture}[
        font={\fontfamily{phv}\selectfont},
        scale=0.52,
        level/.style={thin},
        leveldash/.style={thin,dashed},
        virtual/.style={densely dashed},
        transthick/.style={very thick,-{Stealth[length=2mm,width=2mm]},shorten <=2pt,decoration={snake,amplitude=1.8,post length=1.4mm}},
transthin/.style={thin,-{Stealth[length=1.5mm,width=1.5mm]},shorten <=2pt,decoration={snake,amplitude=1.4,post length=1.4mm}},
        classical/.style={thin,double,<->,shorten >=2pt,shorten <=2pt,>=stealth},
        classical2/.style={thin,double,<->,shorten >=1.5pt,shorten <=1.5pt,>=stealth},
        label/.style={scale=0.8}
        ]
        \def\x{9}
        \def\y{0.25}
        \def\udshift{0.5}
        \node[] at (-1,8) {\textbf{a}};
        \node[] at (7.5,8) {\textbf{b}};
        \node[] at (16,8) {\textbf{c}};
        % for helping shift figure
          \node[scale=0.8] at (-2,-2.0) {};
        % g0 state
        \node[] at (2,0) (g0r) {};
        \node[] at (0,0) (g0l) {};
        \node[] at (1,0) (g0) {};

        \draw[level, thick] (g0r) -- (g0l);
        % g1 state

        % e0 state
        \node[] at (4,3) (e0) {};
        \node[] at (3,3) (e0l) {};
        \node[] at (5,3) (e0r) {};
        \draw[level,thick] (e0r) -- (e0l);
        % e1rabi state
        \node[] at (4,2.4) (e0rab) {};
        \node[] at (5,2.4) (e0rabr) {};
      	\node[] at (3,2.4) (e0rabl) {};
        \draw[leveldash,thick,color=black!50!green] (e0rabl) -- (e0rabr);
        % e1 state
         \node[] at (4,7) (e1) {};
        \node[] at (5,7) (e1r) {};
      	\node[] at (3,7) (e1l) {};
        \draw[level,thick] (e1l) -- (e1r);
        % e1rabi state
        \node[] at (4,6.4) (e1rab) {};
        \node[] at (5,6.4) (e1rabr) {};
      	\node[] at (3,6.4) (e1rabl) {};
        \draw[leveldash,thick,color=black!50!green] (e1rabl) -- (e1rabr);
        
        % e1sbdrive state
        \node[] at (4,5.5) (e1sb) {};
        \node[] at (5,5.5) (e1sbr) {};
      	\node[] at (3,5.5) (e1sbl) {};
        \draw[leveldash,thick,color=black!50!blue] (e1sbl) -- (e1sbr);
        % g0rot state
        \node[] at (1+\x,3.5) (g0rot) {};
        \node[] at (\x,3.5) (g0rotl) {};
        \node[] at (\x+2,3.5) (g0rotr) {};
        \node[] at (\x+\y,3.5) (g0rotlb) {};
        \node[] at (\x+2-\y,3.5) (g0rotrb) {};
        \draw[level,thick,color=black!50!green] (g0rotl) -- (g0rotr);
        % g1rot state
        \node[] at (1+\x,7.0) (g1rot) {};   
        \node[] at (\x,7.0) (g1rotl) {};   
        \node[] at (2+\x,7.0) (g1rotr) {};   
        \node[] at (\x+\y,7.0) (g1rotlb) {};
        \node[] at (\x+2-\y,7.0) (g1rotrb) {};
		\draw[level,thick,color=black!50!green] (g1rotl) -- (g1rotr);
        
                % g1rotud state
        \node[] at (1+\x,7.0-\udshift) (g1rotud) {};   
        \node[] at (\x,7.0) (g1rotudl) {};   
        \node[] at (2+\x,7.0) (g1rotudr) {};   
		\draw[level,thick,color=black!50!green] (g1rotudl) -- (g1rotudr);
         % e0rot state
        \node[] at (4+\x,0.0) (e0rot) {};
        \node[] at (3+\x,0.0) (e0rotl) {};
        \node[] at (5+\x,0.0) (e0rotr) {};
        \node[] at (3 + \y+\x,0.0) (e0rotlb) {};
        \node[] at (5-\y+\x,0.0) (e0rotrb) {};
		\draw[level,thick,color=black!50!green] (e0rotl) -- (e0rotr);
        % e1rot state
        \node[] at (4+\x,3.5) (e1rot) {};
        \node[] at (3+\x,3.5) (e1rotl) {};
        \node[] at (5+\x,3.5) (e1rotr) {};
        \node[] at (3+\y+\x,3.5) (e1rotlb) {};
        \node[] at (5-\y+\x,3.5) (e1rotrb) {};
		\draw[level,thick,color=black!50!green] (e1rotr) -- (e1rotl);

		% state labels
       \node[] at (1,3.5) {$\ket{g1}$};
       \node[] at (1,-0.5) {$\ket{g0}$}; 
       \node[] at (4,7.5) {$\ket{e1}$}; 
       \node[] at (4,3.5) {$\ket{e0}$};
       \node[] at (1+\x-1.5,3.5) {$\ket{\tilde{g}1}$};
       \node[] at (1+3+\x,-0.5) {$\ket{\tilde{g}0}$}; 
       \node[] at (1+\x,7.5) {$\ket{\tilde{e}1}$}; 
       \node[] at (4+\x+1.5,3.5) {$\ket{\tilde{e}0}$};
        
		% Labels and arrows for unrotated frame
               \node[] at (1,4.0) (g1) {}; 
 		\node[] at (2,4.0) (g1r) {};   
        \node[] at (0,4.0) (g1l) {};   
        \draw[level,thick] (g1r) -- (g1l);
		
        \node[color=red] at (-0.5,2) {$\kappa$};  
        \node[color=red] at (6,5) {$\kappa$};  
        \node[] at (1,6) {$\gamma$};  
        \node[] at (4,1.0) {$\gamma$};  
        
        \node[color=red] at (0.5+\x,2) {$\kappa$};  
        \node[color=red] at (4.25+\x,5.25) {$\kappa$};  
        \node[] at (-0.25+\x,5.25) {$\gamma_{+}$};
        \node[] at (2.25+\x,5.25) {$\gamma_{-}$};
        \node[] at (3-0.25+\x,5.25-3.5) {$\gamma_{+}$};
        \node[] at (3+2.25+\x,5.25-3.5) {$\gamma_{-}$};

        \path[transthick,color=red] (g1l) edge[decorate] (g0l);
        \path[transthin,color=black] (e1l) edge[decorate] (g1l);
        \path[transthick,color=red] (e1r) edge[decorate] (e0r);
       	\path[transthin,color=black] (e0r) edge[decorate] (g0r);
        \path[classical,thick,color=black!50!green] (g1) edge[decorate] (e1rab);
       	\path[classical,thick,color=black!50!green] (g0) edge[decorate] (e0rab);
        \path[classical,thick,color=black!50!blue] (g0) edge[decorate] (e1sb);
        \path[classical2,color=black] (5.35,7.15) edge (5.35,6.35);
        \node[scale=0.8,color = black] at (6.0,6.7) {$\Omega_{z}$}  ;
        \node[scale=0.8,color = black!50!green] at (2.25,5.5) {$\Omega_{x}$}  ;
        \node[scale=0.8,color = black!50!blue] at (2.25,3.25) {$\Omega_{b}$}  ;        

        \path[classical2,color=black](5.35,6.5) edge (5.35,5.5);
        \node[scale=0.8,color= black] at (6.0,6.0) {$\Omega_{R}$};
                
        \path[transthin,color=black] (g1rotrb) edge[decorate] (g0rotrb);
        \path[transthin,color=black] (g0rotlb) edge[decorate] (g1rotlb) ;

        \path[transthick,color=red] (g1rotr) edge[decorate] (e1rotr);
       	\path[transthick,color=red] (g0rotl) edge[decorate] (e0rotl);
        \path[transthin,color=black] (e1rotrb) edge[decorate] (e0rotrb);
        \path[transthin,color=black] (e0rotlb) edge[decorate] (e1rotlb) ;

        \draw[-{Latex[length=1mm,width=1mm]},color=black!50!blue,thick,shorten <=1pt] (3+\x,3.5) arc(45:135:0.7cm);
        \draw[-{Latex[length=1mm,width=1mm]},color=black!50!blue,thick,shorten <=1pt] (2+\x,3.5) arc(-135:-45:0.7cm);
        \node[label,color=black!50!blue] at (2.5+\x,4.0) {$g$};
        
%         \draw[-{Latex[length=1mm,width=1mm]},color=black!5!orange,thick,shorten <=1pt] (2+\x/2.0-0.25,3.5) arc(-135:-45:1.0cm);
%         \node[label,color=black!5!orange] at (2.5+\x/2,4.0) {$rf$};
      \end{tikzpicture}
}\hfill
\subfloat{
\begin{tikzpicture}[font={\fontfamily{phv}\selectfont}]
\includegraphics{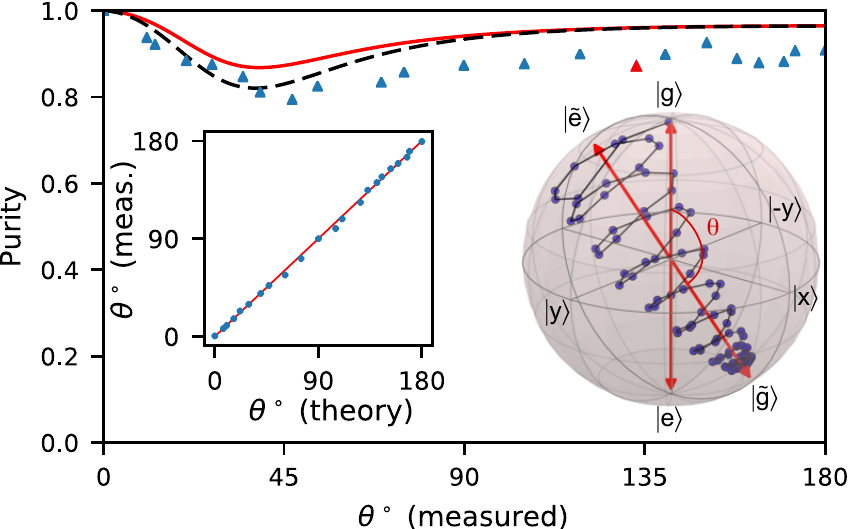}
        \node[] at (8.25,0.0) {};
%         \node[] at (-4,8*0.52) {\textbf{(b)}};
\end{tikzpicture}
}
\caption{Illustration of the universal stabilization scheme for single-qubit states. In the lab frame (a), qubit Rabi drive and blue-sideband modulation are applied with appropriately chosen detuning and strength. In the rotating frame (b), these two drives result in the dressing of the qubit state into arbitrary superpositions $\left|\tilde{g}\right\rangle$, $\left|\tilde{e}\right\rangle$, with resonant coupling between $\left|\tilde{e}0\right\rangle$ to $\left|\tilde{g}1\right\rangle$. Together with the aid of the fast cavity decay, these finally lead to the stabilization of the $\left|\tilde{g}0\right\rangle$ state. (c) The stabilization purity $\left|\left<\vec{\sigma}\right>\right|$, plotted against the polar angle $\theta$ of the stabilization axis, both obtained from qubit tomography. Purities exceeding 80\% are achieved over the entire Bloch sphere, while purities $>$90\% and $>99\%$ are reached for stabilizing the $\left|e\right>$ ($\theta= 180^\circ$) and $\left|g\right>$ ($\theta= 0^\circ$) states, respectively. Experimental data qualitatively agrees with the analytical calculation from Eq.~\eqref{eq:population theory} (red line) and numerical master equation simulation (black dashed line). The stabilization experiment was performed at zero flux, where qubit and cavity frequencies are $\omega_q/2\pi = 4.343$ GHz and $\omega_r/2\pi = 5.439$\,GHz, with the linewidths being $\gamma/2\pi \approx 7.6$\,KHz, $\gamma_{\phi}/2\pi \approx 3$\,KHz and $\kappa/2\pi \approx 1.6$\,MHz.  Left inset: stabilization angles predicted by theory closely match the experimental values. Right inset: trajectory of the qubit state in the dynamic process of stabilization, for the specific case of $\theta = 135^\circ$ (red triangle) with measured purity of 87\%. Starting from $\left|g\right>$, the qubit state moves in a helical path along the stabilization axis, until it saturates around the rotating frame ground state, $\left|\tilde{g}\right>$.}
\label{stabilization}
\end{figure*}
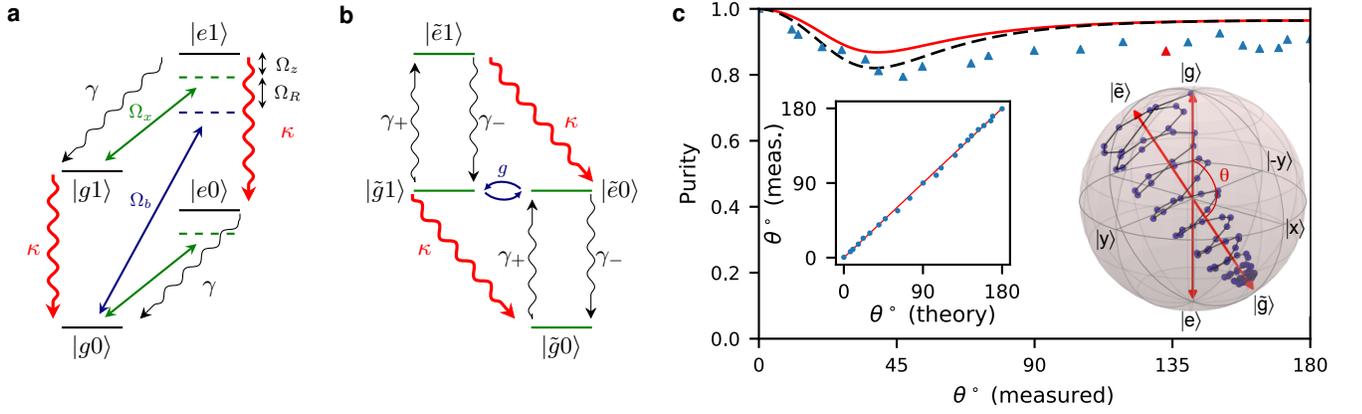

While the red-sideband coupling enables photon-conserving processes, the blue-sideband coupling, which takes place at $\omega_d \approx \omega_{r}'+\omega_{q}'-\chi'$, generates correlated two-photon oscillations between states $\left|g0\right\rangle $ and $\left|e1\right\rangle $.  Interestingly, this interaction produces a much richer resonance structure in transmission (see Fig.~\ref{blue sideband plot}a), which can be accurately reproduced numerically (see Fig.~\ref{blue sideband plot}b). The observed features can be understood conceptually by considering the energy level diagram in the rotating frame, see Fig.~\ref{blue sideband plot}c. The blue-sideband interaction acts as a coherent two-photon pump that drives the circuit to $\left|e1\right\rangle$, causing an avoided crossing between $|g0\rangle$ and $|e1\rangle$ in the level diagram. As the cavity photon loss rate is faster than the qubit decay rate by two orders of magnitude in the experiment ($1/\kappa \approx 100$ ns and $T_1> 20 \mu$s), $\left|e1\right\rangle \rightarrow \left|e0\right\rangle$ is the dominant decay process and traps most of the population in the single-photon subspace in state $\left|e0\right\rangle$. When both photons are eventually lost from the circuit, the state immediately transitions to $\left|e1\right\rangle$, beginning the cycle again. In this sense, the blue-sideband flux drive stabilizes the qubit in the excited state. This, in turn, shifts the cavity frequency down by $2\chi'$ (see B in Fig.~\ref{blue sideband plot}a).  Furthermore, as the blue-sideband interaction splits the degenerate levels of $\left|e1\right\rangle$ and $\left|g0\right\rangle$ in the rotating frame, the cavity transmission measurement actually probes the transitions between $\left|e0\right\rangle$ and $(\left|e1\right\rangle \pm \left|g0\right\rangle)/\sqrt{2}$ so that the avoided crossing is visible within the shifted cavity peak (see C in Fig.~\ref{blue sideband plot}a). Another interesting yet subtle feature is the bright spot observed at the crossing between the un-shifted cavity peak and the avoided crossing (see D in Fig.~\ref{blue sideband plot}a). This corresponds to the scenario where the transition energy between $\left|e0\right\rangle$ and $\left|g0\right\rangle$ in the rotating frame coincides with the energy between $\ket{g0}$ and $\ket{g1}$. As a result, the $\left|g0\right\rangle$ population is replenished weakly by the cavity probe to give rise to an enhanced transmission amplitude at the un-shifted cavity frequency.  

With the blue-sideband coupling being a critical component, we show that it is possible to take a further step towards stabilizing arbitrary states on the Bloch sphere with our tunable coupler circuit. Analogous to coherent population trapping~\cite{Dalton1985CoherentTrapping,Arimondo1996VSpectroscopy} (CPT) but using a harmonic oscillator as the dissipative element, the system is driven with both blue-sideband modulation and qubit Rabi drive at detunings and strengths as shown in Fig.~\ref{stabilization}a. 

Qubit states are dressed by the Rabi drive to become $\left|\tilde{g}\right> = \cos\frac{\theta}{2}\left|g\right>-e^{i\phi}\sin\frac{\theta}{2}\left|e\right>$ and $\left|\tilde{e}\right> = \sin\frac{\theta}{2}\left|g\right>+e^{i\phi}\cos\frac{\theta}{2}\left|e\right>$ in the rotating frame (Fig.~\ref{stabilization}b), where the polar angle $\theta=\arccos \left(\Omega_z/\Omega_R\right)$ is defined by the Rabi drive detuning $\Omega_z$ and the total Rabi frequency $\Omega_R=\sqrt{\Omega_{x}^2+\Omega_z^2}$, while the azimuthal angle $\phi$ determined by the phase of the Rabi drive. The dressing of the qubit states also leads to modified decay and excitation rates between $\left|\tilde{g}\right>$ and $\left|\tilde{e}\right>$ (Fig.~\ref{stabilization}b). These can be found by rewriting the master equation dissipators in the dressed basis as
\begin{align}
\tilde{\gamma}_- & =\gamma \cos^4 \frac{\theta}{2}+\frac{\gamma_{\phi}}{2} \sin^2 \theta,\nonumber \\ 
\tilde{\gamma}_+ & =\gamma \sin^4 \frac{\theta}{2}+\frac{\gamma_{\phi}}{2} \sin^2 \theta,\label{eq:rotating frame decay excitation rate}
\end{align}
where $\gamma$ and $\gamma_{\phi}$ stand for the qubit decay and dephasing rate in zero-temperature lab frame (see Supplementary Information). 
% RN: previous two sentence are a bit long and jumbled IMO. Probably need to be split up

The blue-sideband drive with amplitude $\Omega_b$ provides a resonant interaction of strength $g=\Omega_b \sin^2 \frac{\theta}{2}$ between the rotating frame states $\left|\tilde{g}1\right\rangle$ and $\left|\tilde{e}0\right\rangle$. Along with the fast decay of the resonator, this interaction yields an effective transition rate $\Gamma=4g^2 \kappa/\left(\kappa^2+4g^2\right)$ among qubit states $\left|\tilde{e}\right\rangle$ and $\left|\tilde{g}\right\rangle$. This produces an overall qubit decay rate of $\tilde{\gamma}_{-}+\Gamma$ that competes against the excitation rate $\tilde{\gamma}_{+}$, to stabilize the effective ground state $\left|\tilde{g}\right>$ with a population of
\begin{equation}
P_{\tilde{g}}=\frac{\gamma_{-}+\Gamma}{\gamma_{-}+\gamma_{+}+\Gamma}.\label{eq:population theory}
\end{equation}
As both polar and azimuthal angles of $\left|\tilde{g}\right>$ can be easily manipulated in the experiment, this scheme allows for stabilization along an arbitrary direction with high fidelity. 

We apply this protocol to demonstrate stabilization of arbitrary states on the Bloch sphere. The polar angle was varied by changing the Rabi drive detuning $\Omega_z$ while keeping its strength $\Omega_x/2\pi$ fixed at 9\,MHz. As can be seen from Eq.~\eqref{eq:population theory}, the azimuthal angle has no effect on the stabilization fidelity and was thus set to zero. The amplitude of the flux modulation is calibrated to create a constant blue-sideband coupling strength $\Omega_b/2\pi = 0.5$\,MHz for all stabilization angles, with the detuning chosen in each case to be $\Omega_{z} + \Omega_{R}$. The measured stabilization purity $\left|\left<\vec{\sigma}\right>\right| = \sqrt{\left<\sigma_x\right>^2+\left<\sigma_y\right>^2+\left<\sigma_z\right>^2}$ is plotted as a function of the stabilization polar angle $\theta$ in Fig.~\ref{stabilization}c, which closely follow the theory prediction made by Eq.~\eqref{eq:population theory}. The excited state $\left|e\right>$ is stabilized with 93$\%$ purity at $\theta = 180^\circ$, where only flux modulation at the blue-sideband frequency is needed. Purity starts to reduce as $\theta$ is lowered, which can be understood by the blue-sideband interaction losing efficiency in coupling the $\left|\tilde{g}1\right>$ and $\left|\tilde{e}0\right>$ states when the rotating-frame ground state $\left|\tilde{g}\right>$ has less overlap with the bare excited state, $\left|e\right>$. This, however, does not invalidate the scheme's performance for small angles. According to Eq.~\eqref{eq:rotating frame decay excitation rate}, the qubit's natural decay guarantees $\tilde{\gamma}_{-} \gg \tilde{\gamma}_+$ as $\theta \rightarrow 0$, resulting in good stabilization fidelity in Eq.~\eqref{eq:population theory}, irrespective of how small $\Gamma$ is. This is reflected in Fig.~\ref{stabilization}c as a revival of the purity from a minimum value of $\sim80\%$ to near unity (limited by lab-frame qubit temperature) at $\theta = 0$, where the lab-frame ground state $\left|g\right>$ is ``stabilized'' through the natural decay of the qubit. The high fidelity at all stabilization angles therefore relies upon the mixed contribution of the active stabilization process induced by the blue-sideband interaction ($\Gamma$), and the passive process from natural qubit decay ($\tilde{\gamma}_{-}$).

In summary, we have demonstrated a cavity-assisted, autonomous protocol for universal qubit state stabilization, an important step towards stabilization of many-body states~\cite{Ma2017AutonomousSolids,Anderson2016EngineeringArrays} and autonomous error correction~\cite{Leghtas2013Hardware-EfficientProtection,Kapit2015PassiveFabric, Kapit2016Hardware-EfficientCircuits}.  The circuit developed in this work provides a flux-controlled tunable coupling between two fixed frequency modes, and maintains excellent coherence over the majority of the tuning range. In addition to stabilization, the circuit is capable of producing red-sideband interactions, which are critical for frequency conversion, random access gates and quantum communication.  Finally, a single tunable coupler can support several modes, significantly reducing the complexity of large quantum circuits and their associated room-temperature electronics.

\begin{acknowledgments} 
We thank M. W. Wei, Andy C. Y. Li and J. Lawrence for helpful discussions.
This material is based upon work supported by the Army Research Office under (W911NF-15-2-0058) and DOD contract H98230-15-C0453. Use of the Center for Nanoscale Materials, an Office of Science user facility, was supported by the U. S. Department of Energy, Office of Science, Office of Basic Energy Sciences, under Contract No. DE-AC02-06CH11357.  This work made use of the Pritzker Nanofabrication Facility of the Institute for Molecular Engineering at the University of Chicago, which receives support from SHyNE, a node of the National Science Foundation's National Nanotechnology Coordinated Infrastructure (NSF NNCI-1542205). E. Kapit was supported by the Louisiana Board of Regents grant (LEQSF(2016-19)-RD-A-19. We gratefully acknowledge support from the David and Lucile Packard Foundation.
\end{acknowledgments} 

\bibliography{tunable_coupler_noissn}

\onecolumngrid
\newpage
\appendix
\renewcommand{\thefigure}{S\arabic{figure}}
\setcounter{figure}{0}
\renewcommand\theequation{S\arabic{equation}}
\setcounter{equation}{0}

\section{Sample fabrication and Experimental setup}
\begin{figure}[h]
\includegraphics[scale=0.8]{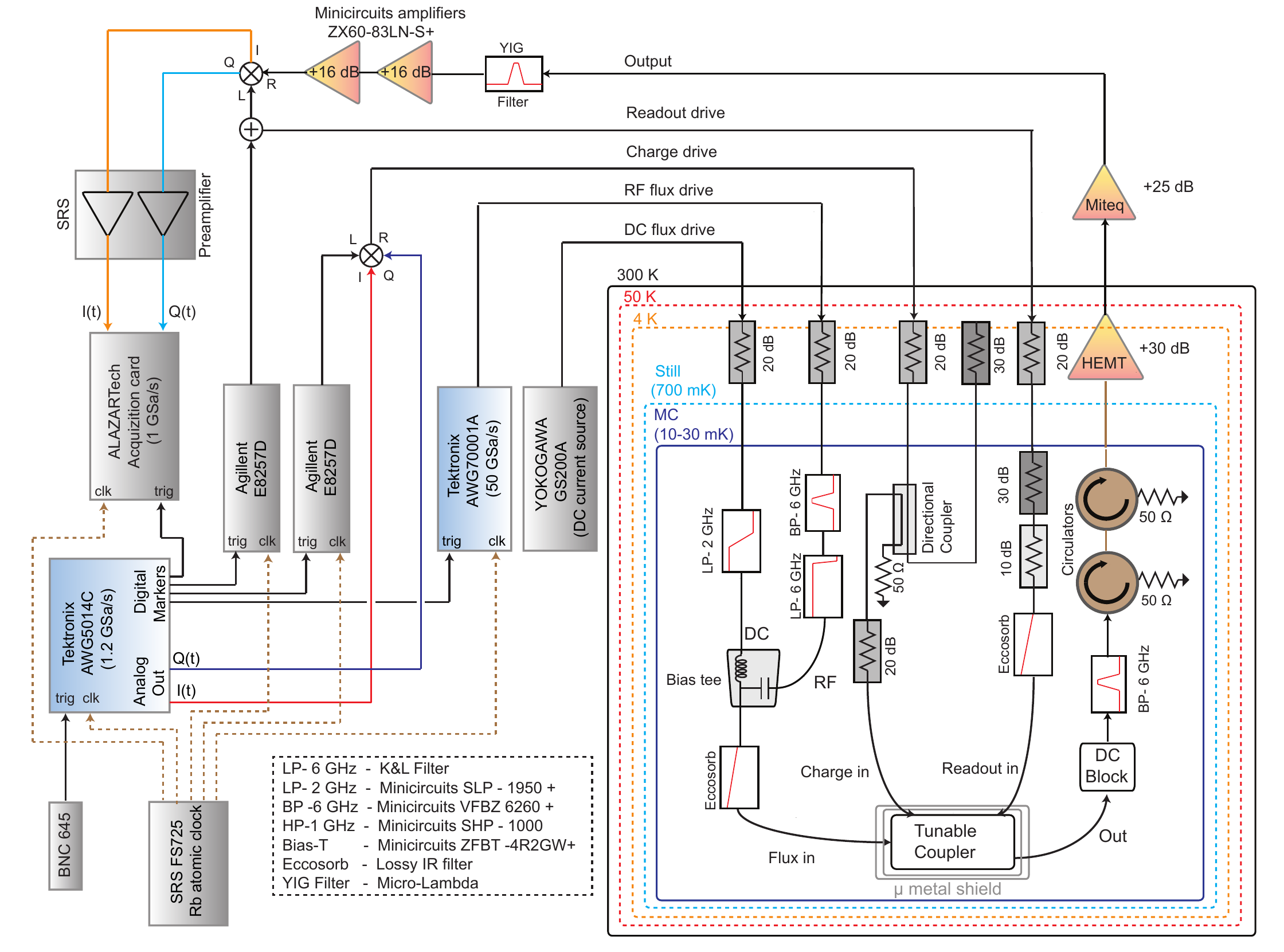}
\caption{Schematic of the experimental setup, including the cryogenic and room-temperature control instrumentation.
\label{wiring diagram}}
\end{figure}

The base layer of the sample is made of 150\,nm of niobium sputtered on 430 \,$\mu$m thick C-plane sapphire substrate, which is then patterned by optical lithography and reactive ion etching (RIE) to define the optical part of the circuit (excluding the qubit and the coupler part). The qubit and the coupler junctions are both fabricated with Manhattan pattern, with the bottom (80\,nm) and the top (150\,nm) aluminum layers deposited via dual-angle electron-beam evaporation. The two layers are gapped by an Al$_x$O$_y$ insulator layer grown in an oxidation process under 20\,mBar of high-purity $\rm O_2$ for 12 minutes. The size of the coupler junctions is designed to be 100 times larger than the qubit junction, which guarantees that $L_{g0}\ll L_{r,\,}L_{q}$ and allows the adiabatic elimination of the coupler mode. We make sure the SQUID loop size is fabricated small enough so that the geometric inductance of loop itself does not become comparable to the Josephson inductance of the SQUID junctions, in order to avoid undesired hysteresis.

The schematic of the instrumentation and cryogenic setup can be seen in Fig.~\ref{wiring diagram}. The device is mounted and wirebonded to a multilayer copper PCB (IBM-type) microwave-launcher board, which is then heat sunk to the base stage of a Bluefors dilution refrigerator (10-30 mK) via an OFHC copper post. The sample is surrounded by a can containing two layers of $\mu$-metal shielding, thermally anchored using an inner close fit copper shim sheet, attached to the copper can lid. The device is connected to the rest of the setup through four ports: a charge port that applies qubit drive tones, an input and an output port for readout drive tones, a flux port for shifting the qubit frequency using a dc-flux bias current and for applying rf sideband flux pulses. The charge pulses are generated by mixing a local oscillator tone (generated from an Agilent 8257D rf signal generator), with pulses generated by a Tektronix AWG5014C arbitrary waveform generator (TEK) with a sampling rate of 1.2 GSa/s, using an IQ-Mixer (MARQI MLIQ0218).  The readout drive pulse is generated from a second Agilent 8257D rf signal generator, which is also controlled by digital trigger pulses from the TEK. The flux-modulation pulses are directly synthesized by a Tektronix AWG70001A arbitrary waveform generator (50 GSa/s) and attenuated by $20$ dB at the 4 K stage. Filters on the rf flux line are configured to create a pass band between 4.8 GHz - 6 GHz, which allows blue-sideband modulation while cutting off noises at the qubit frequency. For red-sideband flux modulation, a Low pass filter (Minicircuits VLF -1800+) at 2 GHz is used instead. A better filtering option for the simultaneous implementation of both sidebands could be using a notch (band stop) filter, with a rejection band covering only the qubit frequency and allowing both the red and blue frequency to pass through. The dc flux bias current is generated by a YOKOGAWA GS200 low-noise current source, attenuated by 20 dB at the 4 K stage, and low-pass filtered down to a bandwidth of 2 MHz. The dc flux bias current is combined with the flux-modulation pulses at a bias tee thermalized at the base stage. The state of the transmon is measured using the transmission of the readout resonator, through the dispersive circuit QED readout scheme. The transmitted signal from the readout resonator is passed through a set of cryogenic circulators (thermalized at the base stage) and amplified using a HEMT amplifier (thermalized at the 4 K stage). Once out of the fridge, the signal is filtered (tunable narrow band YIG filter with a bandwidth of 80 MHz) and further amplified. The amplitude and phase of the resonator transmission signal are obtained through a homodyne measurement, with the transmitted signal demodulated using an IQ mixer and a local oscillator at the readout resonator frequency. The homodyne signal is amplified (SRS preamplifier) and recorded using a fast ADC card (ALAZARtech).

\section{Circuit quantization with linear model}

\begin{figure}
\includegraphics[scale=0.4]{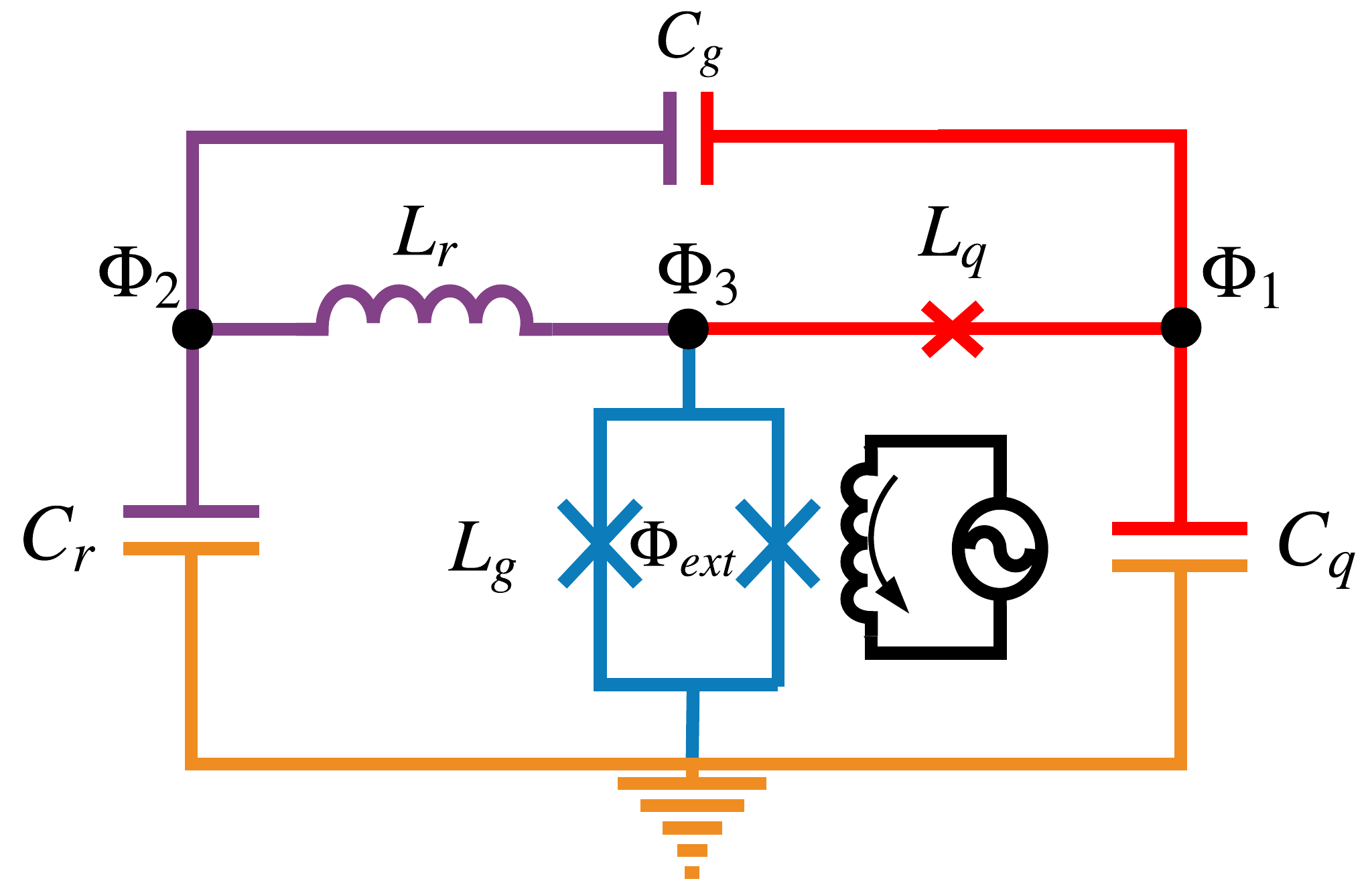}
\caption{Circuit schematic of the tunable coupling device. \label{circuit diagram}}
\end{figure}

We begin by linearizing the circuit shown in Fig.~\ref{circuit diagram}, where the non-linear inductive components,
the transmon qubit junction and the SQUID, are simplified as linear
inductors $L_{j}$ and $L_{g}$. The linear inductance of the SQUID
is tunable with flux $L_{g}=\frac{L_{g0}}{\left|\cos(\Phi_{ext}/2\Phi_{0})\right|}$.
We denote the node flux variables $\Phi_{1}$, $\Phi_{2}$ and $\Phi_{3}$,
and circuit Lagrangian is given by

\begin{equation}
\mathcal{L}=-\frac{\left(\Phi_{1}-\Phi_{3}\right)^{2}}{2L_{q}}-\frac{\left(\Phi_{2}-\Phi_{3}\right)^{2}}{2L_{r}}-\frac{\Phi_{3}^{2}}{2L_{g}}+\frac{C_{q}\dot{\Phi}_{1}^{2}}{2}+\frac{C_{r}\dot{\Phi}_{2}^{2}}{2}+\frac{C_{g}\left(\dot{\Phi}_{1}-\dot{\Phi}_{2}\right)^{2}}{2}.
\end{equation}
Charge variables conjugate to the flux can be found from a Legendre
transformation

\begin{equation}
Q_{i}=\frac{\partial\mathcal{L}}{\partial\dot{\Phi}_{i}},
\end{equation}
and the circuit Hamiltonian can be obtained via

\begin{align}
\mathcal{H} & =\dot{\Phi}_{i}Q_{i}-\mathcal{L}\\
 & =\frac{\left(\Phi_{1}-\Phi_{3}\right)^{2}}{2L_{q}}+\frac{\left(\Phi_{2}-\Phi_{3}\right)^{2}}{2L_{r}}+\frac{\Phi_{3}^{2}}{2L_{g}}+\frac{1}{2C_{*}^{2}}\left[\left(C_{r}+C_{g}\right)Q_{1}^{2}+\left(C_{q}+C_{g}\right)Q_{2}^{2}+2C_{g}Q_{1}Q_{2}\right],
\end{align}
where 

\begin{equation}
C_{*}^{2}=C_{r}C_{q}+C_{r}C_{g}+C_{q}C_{g}.
\end{equation}
Obviously $\Phi_{3}$ is a free degree of freedom, which can be eliminated from minimizing the Hamiltonian,

\begin{equation}
\frac{\partial\mathcal{H}}{\partial\Phi_{3}}=0,
\end{equation}
which gives

\begin{equation}
\Phi_{3}=\frac{L_{g}}{L_{*}^{2}}\left(L_{r}\Phi_{1}+L_{q}\Phi_{2}\right),
\end{equation}
where

\begin{equation}
L_{*}^{2}=L_{r}L_{q}+L_{r}L_{g}+L_{q}L_{g}.
\end{equation}
This circuit is then described by the following two-body Hamiltonian
that has both capacitive and inductive coupling terms,

\begin{equation}
\mathcal{H}=\frac{\Phi_{1}^{2}}{2L_{q}}\left(1-\frac{L_{r}L_{g}}{L_{*}^{2}}\right)+\frac{\Phi_{2}^{2}}{2L_{r}}\left(1-\frac{L_{q}L_{g}}{L_{*}^{2}}\right)-\frac{L_{g}}{L_{*}^{2}}\Phi_{1}\Phi_{2}+\frac{1}{2C_{*}^{2}}\left[\left(C_{r}+C_{g}\right)Q_{1}^{2}+\left(C_{s}+C_{g}\right)Q_{2}^{2}+2C_{g}Q_{1}Q_{2}\right].
\end{equation}

With the flux and charge operators expressed in terms of creation and annihilation
operators

\begin{align}
\Phi_{i} & =\sqrt{\frac{\hbar Z_{i}}{2}}\left(a_{i}^{\dagger}+a_{i}\right),\\
Q_{i} & =i\sqrt{\frac{\hbar}{2Z_{i}}}\left(a_{i}^{\dagger}-a_{i}\right)
\end{align}
where

\begin{equation}
Z_{1}=\sqrt{\frac{L_{q}\left(C_{r}+C_{g}\right)}{C_{*}^{2}\left(1-\frac{L_{r}L_{g}}{L_{*}^{2}}\right)}},
\end{equation}

\begin{equation}
Z_{2}=\sqrt{\frac{L_{r}\left(C_{q}+C_{g}\right)}{C_{*}^{2}\left(1-\frac{L_{j}L_{g}}{L_{*}^{2}}\right)}},
\end{equation}
the Hamiltonian is rewritten as

\begin{equation}
H=\hbar\omega_{q}a_{1}^{\dagger}a_{1}+\hbar\omega_{r}a_{2}^{\dagger}a_{2}+g_{L}\left(a_{1}^{\dagger}+a_{1}\right)\left(a_{2}^{\dagger}+a_{2}\right)+g_{C}\left(a_{1}^{\dagger}-a_{1}\right)\left(a_{2}^{\dagger}-a_{2}\right),\label{Eq:Circuit Hamiltonian Appendix}
\end{equation}
where

\begin{equation}
\omega_{q}=\sqrt{\frac{\left(1-\frac{L_{r}L_{g}}{L_{*}^{2}}\right)\left(C_{r}+C_{g}\right)}{L_{q}C_{*}^{2}}},
\end{equation}

\begin{equation}
\omega_{r}=\sqrt{\frac{\left(1-\frac{L_{q}L_{g}}{L_{*}^{2}}\right)\left(C_{q}+C_{g}\right)}{L_{r}C_{*}^{2}}},
\end{equation}

\begin{equation}
g_{L}=-\frac{\hbar L_{g}}{2L_{*}^{2}}\sqrt{Z_{1}Z_{2}},
\end{equation}

\begin{equation}
g_{C}=-\frac{\hbar C_{g}}{2C_{*}^{2}}\sqrt{\frac{1}{Z_{1}Z_{2}}}.
\end{equation}

By rearranging the interaction terms in Eq.~\eqref{Eq:Circuit Hamiltonian Appendix}, red and blue coupling given by Eq.~\eqref{eq:grb} in the main text can be retrieved.

\section{The calibration of the $g_r$ coupling strength}

To calibrate the static coupling strength of $g_r$ as a function of the flux, two methods have been employed in the experiment. The first one is to make use of the photon number splitting of the qubit peak that can be observed from the two-tone measurement of the qubit spectroscopy, shown in Fig.~\ref{spec_and_coherence}a of the main text. $g_r$ can thus be directly calculated using the formula
\begin{equation}
\left|g_r\right|=\sqrt{\frac{2\chi \Delta (\Delta+\alpha)}{\alpha}},
\end{equation}
where both the anharmonicity $\alpha$ and the qubit-cavity detuning $\Delta$ are easily obtained from spectroscopy measurements. 

At flux values where the coupling strength is not strong enough to resolve the number splitting, we take a different approach by applying a voltage drive with strength $\epsilon_d$ on the cavity at the qubit frequency, and measuring the Rabi rate of the qubit,

\begin{equation}
\Omega_R=2\epsilon_d\left|\frac{g_r}{\Delta}\right|+\Omega_0,\label{Eq:Rabi rate}
\end{equation}
where the first term represents the perturbative strength of the cavity drive on the qubit, and the second term, which is a constant rate, is due to the spurious coupling between the cavity drive line to the qubit capacitor pad. $\epsilon_d$ and $\Omega_0$ can be calibrated by fitting Eq.~\eqref{Eq:Rabi rate} with $g_r/\Delta$ and $\Omega_R$ measurement values (taken in the same flux range where the number-splitting is still well resolved). With calibrated $\epsilon_d$ and $\Omega_0$,  Eq.~\eqref{Eq:Rabi rate} is capable of providing $g_r$ across the entire flux range.

\section{A general scheme for the stabilization of single-qubit state}

In this appendix we demonstrate a theoretical scheme for stabilizing arbitrary single-qubit state, through a qubit-cavity Hamitonian of the form
\begin{equation}
H=H_q+H_{\textup{int}}+H_c,\label{eq:rotating frame Hamiltonian}
\end{equation}
where the qubit term $H_q=\frac{\Omega_R}{2}\vec{r}\cdot\vec{\sigma}$ is a spin-$\frac{1}{2}$ Hamiltonian subject to a magnetic field $\vec{B}=\Omega_R\vec{r}$, and $H_c=\delta a^\dagger a$ represents a lossy cavity that is coupled via some interaction $H_{\textup{int}}$ to the qubit. We assume this is a rotating frame Hamiltonian resulted by some external drives, without worrying for now about its realization. 

For simplicity and w.l.o.g we choose $\vec{B}$ to have only $\vec{z}$ and $\vec{x}$ components, which lets us write

\begin{equation}
H_q=\frac{1}{2} (\Omega_x \sigma_x+\Omega_z \sigma_z)=\frac{\Omega_R}{2}(\sigma_x\sin\theta+\sigma_z\cos\theta),
\end{equation}
where $\Omega_R=\sqrt{\Omega_x^2+\Omega_z^2}$ is the qubit's total Rabi frequency, and $\theta=\arccos(\Omega_z/\Omega_R)$. The rotation matrix $U$

\begin{equation}
U=\begin{pmatrix}
\cos\frac{\theta}{2}&\sin\frac{\theta}{2}\\ 
-\sin\frac{\theta}{2}&\cos\frac{\theta}{2}
\end{pmatrix}\label{eq:unitary transformation}
\end{equation}
connects the rotating frame eigenstates of the qubit to its lab frame basis,

\begin{equation}
\left|\tilde{g}\right> = U \left|g\right>=\cos\frac{\theta}{2}\left|g\right>-\sin\frac{\theta}{2}\left|e\right>,\label{eq:rotating frame ground state}
\end{equation}
\begin{equation}
\left|\tilde{e}\right> = U \left|e\right>=\sin\frac{\theta}{2}\left|g\right>+\cos\frac{\theta}{2}\left|e\right>.\label{eq:rotating frame excited state}
\end{equation}

\begin{figure}

\centering
\includegraphics[scale=0.4]{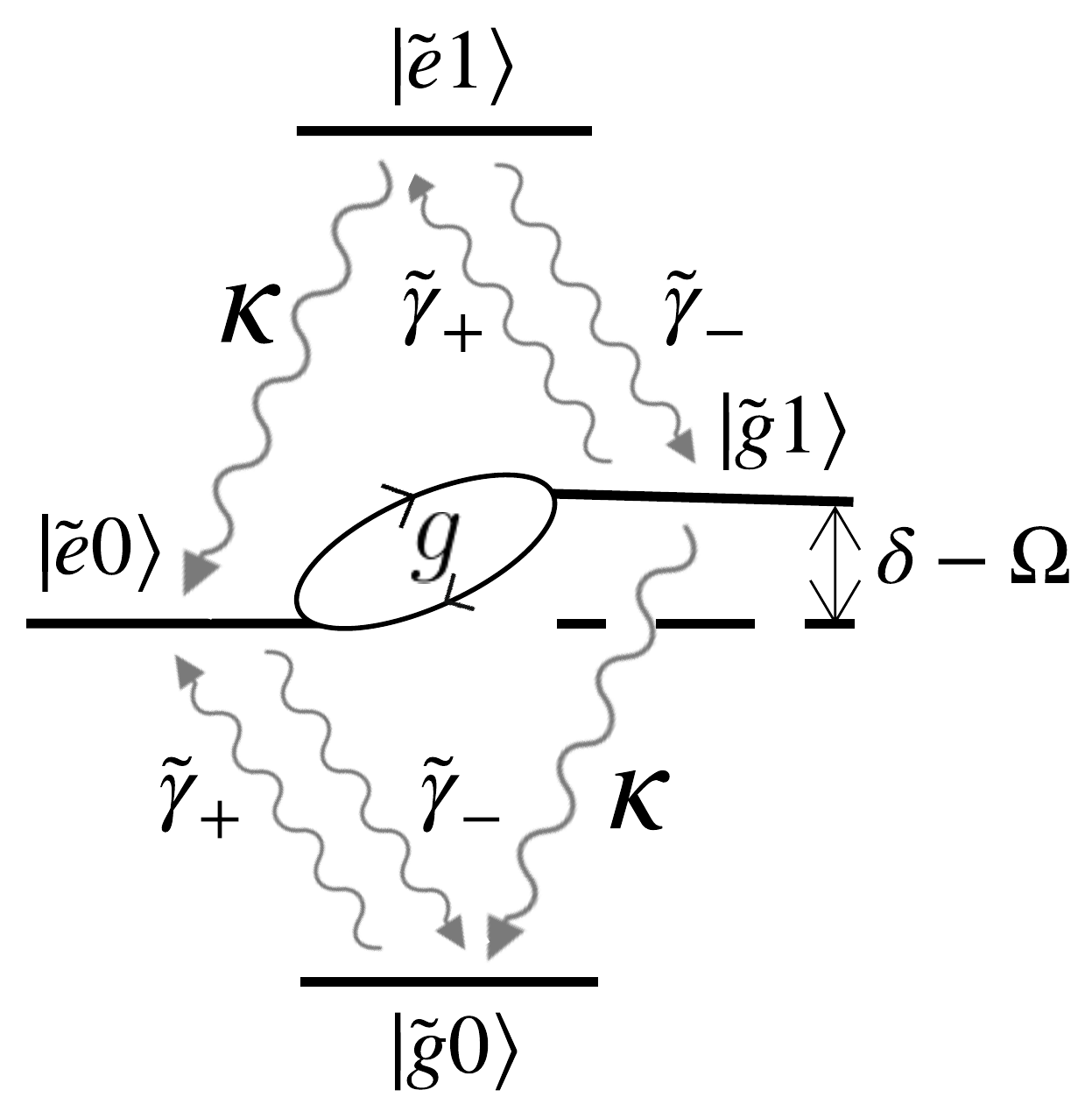}
 
\caption{The decay diagram consisting the lowest four Floquet levels. The lab frame decay rates translate to these rotating frame decay rates via a unitary transformation acting on the lab frame dissipators. $\left| \tilde{e} 0 \right>$ and $\left| \tilde{g} 1 \right>$ coupled with strength $g$ can be brought to resonance by tuning the qubit or cavity frequency, while all other levels are far detuned.}

\label{fig:four level diagram}

\end{figure}

Here and throughout the text, objects with tilde stand for those in the rotating frame. The decay and excitation rate between $\left|\tilde{g}\right>$ and $\left|\tilde{e}\right>$ can be easily calculated by rewriting the lab frame dissipators (at zero temperature) in the new basis, 

\begin{equation}
\gamma \mathcal{D}[\sigma^-]\rho =\gamma \mathcal{D}[U^\dagger \tilde{\sigma}^- U]\rho= \gamma \mathcal{D}\left[\frac{\tilde{\sigma}_z}{2}\sin \theta -\tilde{\sigma}^+\sin^2 \frac{\theta}{2}+\tilde{\sigma}^-\cos^2 \frac{\theta}{2}\right]\rho,\label{eq:lab frame decay}
\end{equation}
\begin{equation}
\gamma_{\phi} \mathcal{D}[\sigma_z]\rho =\gamma_{\phi} \mathcal{D}[U^\dagger \tilde{\sigma}_z U]\rho= \gamma_{\phi} \mathcal{D}\left[\tilde{\sigma}_z\cos \theta -(\tilde{\sigma}^+ + \tilde{\sigma}^-) \sin\theta\right]\rho,\label{eq:lab frame dephasing}
\end{equation}
where $\gamma$ and $\gamma_{\phi}$ are the decay and dephasing rate of the qubit in the lab frame. Therefore, by regrouping the above dissipators and dropping out the fast oscillating terms (assuming $\Omega_R \gg \gamma,\gamma_{\phi}$), such as $\tilde{\sigma}^+ \rho \tilde{\sigma}^+$ and $\tilde{\sigma}^- \rho \tilde{\sigma}^-$ etc., we obtain the effective decay rate $\tilde{\gamma}_-$, excitation rate $\tilde{\gamma}_+$ and dephasing rate $\tilde{\gamma}_{\phi}$ in the rotating frame as

\begin{equation}
\tilde{\gamma}_-=\gamma \cos^4 \frac{\theta}{2}+\frac{\gamma_{\phi}}{2} \sin^2 \theta,\label{eq:rotating frame decay rate}
\end{equation}

\begin{equation}
\tilde{\gamma}_+=\gamma \sin^4 \frac{\theta}{2}+\frac{\gamma_{\phi}}{2} \sin^2 \theta,\label{eq:rotating frame excitation rate}
\end{equation}

\begin{equation}
\tilde{\gamma}_{\phi}=\frac{\gamma}{2} \sin^2 \theta+\gamma_{\phi} \cos^2 \theta.\label{eq:rotating frame dephasing rate}
\end{equation}

Now we can write down the master equation in the rotating frame basis,

\begin{equation}
\dot{\rho}=-i[H,\rho]+\kappa \mathcal{D}[a]\rho + \tilde{\gamma}_- \mathcal{D}[\tilde{\sigma}_-]\rho + \tilde{\gamma}_+ \mathcal{D}[\tilde{\sigma}_+]\rho + \frac{\tilde{\gamma}_{\phi}}{2} \mathcal{D}[\tilde{\sigma}_z]\rho,\label{eq:master equation}
\end{equation}
where $\kappa$ is the cavity's photon loss rate. Our goal is to stabilize the qubit in its rotating frame ground state $\left|\tilde{g}\right>$. To gain more insight into the stabilization process, we focus on the dynamics of the lowest four energy levels of Eq.~\eqref{eq:rotating frame Hamiltonian} (which is well justified when the energy scale of the rotating frame Hamiltonian is small compared to the anharmonicity of the qubit), illustrated by Fig.~\ref{fig:four level diagram}. Without coupling to the cavity, the ratio of the excitation rate and the decay rate sets the ``rotating frame temperature" $\tilde{T}$ of the qubit

\begin{equation}
\frac{\tilde{\gamma}_+}{\tilde{\gamma}_-}=e^{-\frac{\hbar \Omega}{k_B \tilde{T}}},
\end{equation}
which further sets the qubit's population distribution. However, when $\left| \tilde{e} 0 \right>$ and $\left| \tilde{g} 1 \right>$ are coupled together through $H_{\textup{int}}$ with strength $g$,

\begin{equation}
g=\left< \tilde{e} 0 \right| H_{\textup{int}} \left| \tilde{g} 1 \right>,\label{eq:g}
\end{equation}
the qubit can lose its excitation and scatter a Raman photon in the cavity mode, which is again lost through the cavity decay channel that brings $\left| \tilde{g} 1 \right>$ back to $\left| \tilde{g} 0 \right>$, and autonomously completes the stabilization process. The $\left| \tilde{e} 0 \right> \rightarrow \left| \tilde{g} 1 \right> \rightarrow \left| \tilde{g} 0 \right>$ transition can be thought of as a cavity assisted qubit decay channel, which is sometimes referred to as the "refilling" process~\cite{Kapit2014InducedLight}. Intuitively, the success of the scheme with high stabilization fidelity lies upon $\kappa \gg \tilde{\gamma}_+$ as well as a decent $\left| \tilde{e} 0 \right> \rightarrow \left| \tilde{g} 1 \right>$ transition rate $\Gamma$.

\begin{figure}

\centering
\includegraphics[scale=0.4]{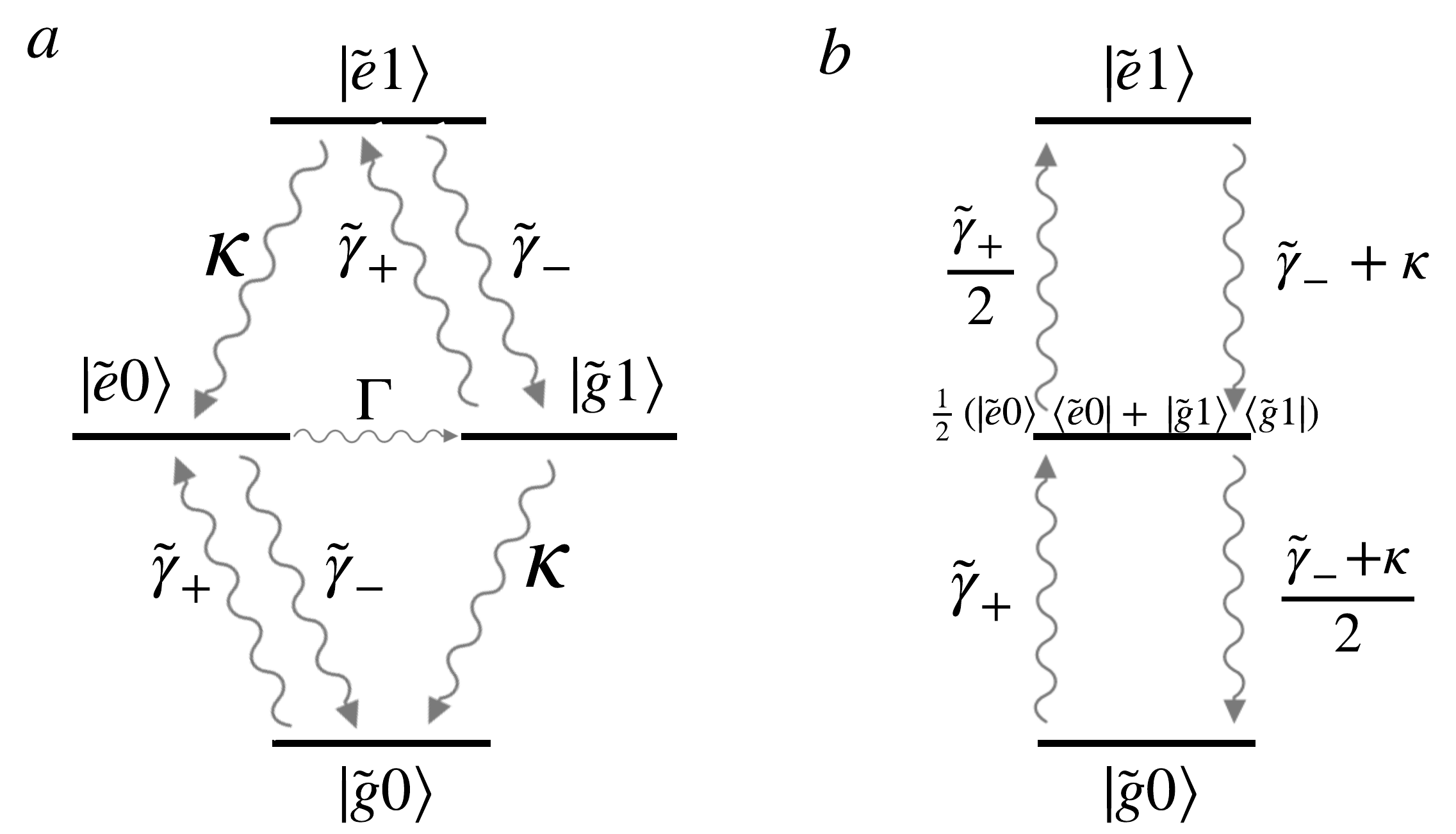}
 
\caption{Decay diagrams for (a) weak and (b) strong coupling regimes, depending on $g/\kappa$. We set $\delta=\Omega_R$ for both cases for optimal performance. (a) When $g/\kappa \ll 1$, the $\left| \tilde{e} 0 \right>$ state can decay back to the ground state $\left| \tilde{g} 0 \right>$ through a two-step process $\left| \tilde{e} 0 \right> \rightarrow \left| \tilde{g} 1 \right> \rightarrow \left| \tilde{g} 0 \right>$, which is limited by the slower rate of the two. $\Gamma$ is calculated from Fermi's golden rule as $4g^2/\kappa$. (b) When $g/\kappa \gg 1$, the transition rate is strong enough to build up population in $\left| \tilde{g} 1 \right>$ and even allow the photon to tunnel back to $\left| \tilde{e} 0 \right>$ before it is lost, giving rise to a coherent oscillation between the two levels fast enough to be viewed as a equally weighted mixture. As the photon spends half of its time in each mode, its decay rate to $\left| \tilde{g} 0 \right>$ is effectively the average of $\kappa$ and $\gamma$, and it jumps to $\left| \tilde{e} 1 \right>$ at half of the excitation rate $\tilde{\gamma}_+$, as this transition is only allowed when the photon lives in the qubit mode. Similarly, we can find the rest of the decay rates for this approximate three level system. Finally, by solving the optical Bloch equations we arrive to the analytical expressions of the stabilization fidelity, given by Eq.~\eqref{eq:weak coupling fidelity} and Eq.~\eqref{eq:strong coupling fidelity}.}

\label{fig:weak and strong coupling regimes}

\end{figure}

To begin our treatment with a more quantitative analysis, under the assumption that the cavity decay rate is dominant among all dissipation rates, we divide the parameter space into two different regimes in terms of the ratio $g/\kappa$, namely the weak coupling regime ($g/\kappa \ll 1$) and the strong coupling regime ($g/\kappa \gg 1$), as shown in Fig.~\ref{fig:weak and strong coupling regimes}. In the weak coupling regime, the $\left| \tilde{g} 1 \right>$ state can not build up any population as the photon is very quickly drained. Therefore $\left| \tilde{e} 0 \right>$ exponentially decays at the transition rate $\Gamma$ given by Fermi's golden rule ~\cite{Clerk2010IntroductionAmplification}

\begin{equation}
\Gamma=\frac{g^2\kappa}{(\kappa/2)^2+(\delta-\Omega)^2},
\end{equation}
where the transition rate is maximized at $\delta=\Omega$ and reduces simply to

\begin{equation}
\Gamma=\frac{4g^2}{\kappa}.
\end{equation}
In the weak coupling regime, the population of $\left| \tilde{e} 0 \right>$ varies at a rate much slower than $\kappa$. Thus the qubit dissipation terms associated with $\tilde{\gamma}_{\pm}$ can be linearly added into the optical Bloch equations. Those can be straightforwardly solved for the qubit ground state population as,

\begin{equation}
P_{\tilde{g}}=P_{\tilde{g}0}+P_{\tilde{g}1}=\frac{\tilde{\gamma}_- (\tilde{\gamma}_-+\tilde{\gamma}_++\kappa)\kappa^2 +4 g^2 (\tilde{\gamma}_-+\kappa)(\tilde{\gamma}_++\kappa)}{(\tilde{\gamma}_-+\tilde{\gamma}_+)(\tilde{\gamma}_-+\tilde{\gamma}_++\kappa)\kappa^2 +4g^2 [(\tilde{\gamma}_-+\tilde{\gamma}_+)(\tilde{\gamma}_++\kappa)+\kappa^2]}.\label{eq:weak coupling fidelity}
\end{equation}

% \begin{figure}
% \begin{tikzpicture}
% \node[anchor=north west] (image) at (-1.5,0) {\includegraphics[width=.21\textwidth]{images/pg_vs_g.png}};
% \node[anchor=north west] (image) at (-1.5,4) {\includegraphics[width=.21\textwidth]{images/p1_vs_g.png}};
% \end{tikzpicture}
% \caption{(a) The stabilized state population and (b) mean cavity photon number for stabilization angle $\theta=\pi$, as a function of the coupling strength $g=\left< \tilde{e} 0 \right| H_{int} \left| \tilde{g} 1 \right>$, calculated with $\kappa/2\pi$= 1\,MHz, $\gamma/2\pi$ = 0.1\,MHz, $\gamma_{\phi}/2\pi$ = 0.1\,MHz. Parameters are exaggerated for enhancing visual contrast and do not reflect experimental values. At $\theta = \pi$, the rotating frame ground state overlaps with the lab frame excited state, i.e. $\left| \tilde{g} \right>=\left| e \right>$. The exact solution from the master equation coincides with the weak coupling formula when $g$ is small, whereas it falls into its asymptote predicted by the strong coupling formula when $g$ is big, showing an good agreement between the theory and the numerical calculation. \label{fig:comparison}}

% \end{figure}

\begin{figure}
\subfloat{
  \includegraphics[width=0.5\textwidth]{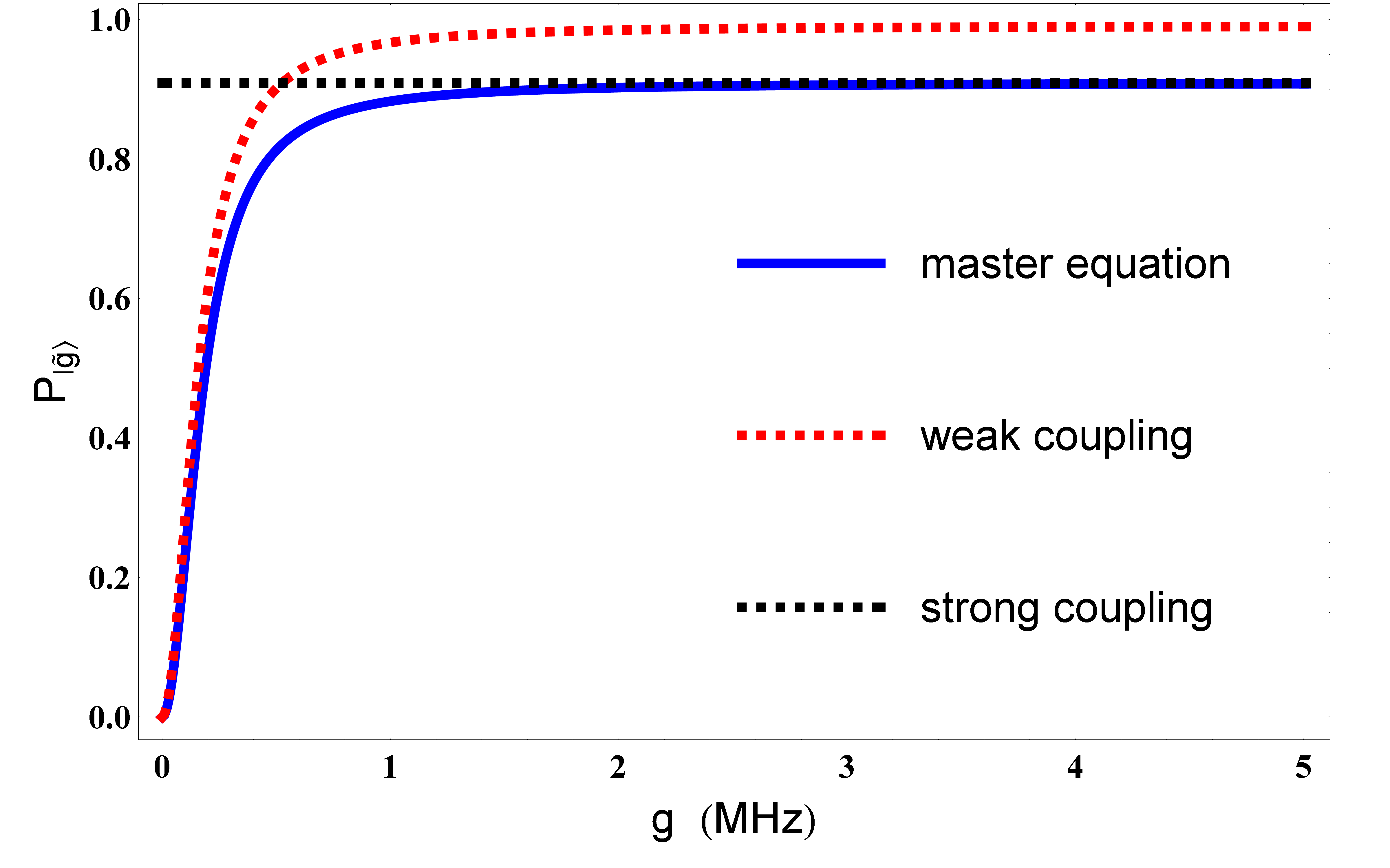}
  \label{fig:fidelity comparison}
  }
  \subfloat{
  \includegraphics[width=0.5\textwidth]{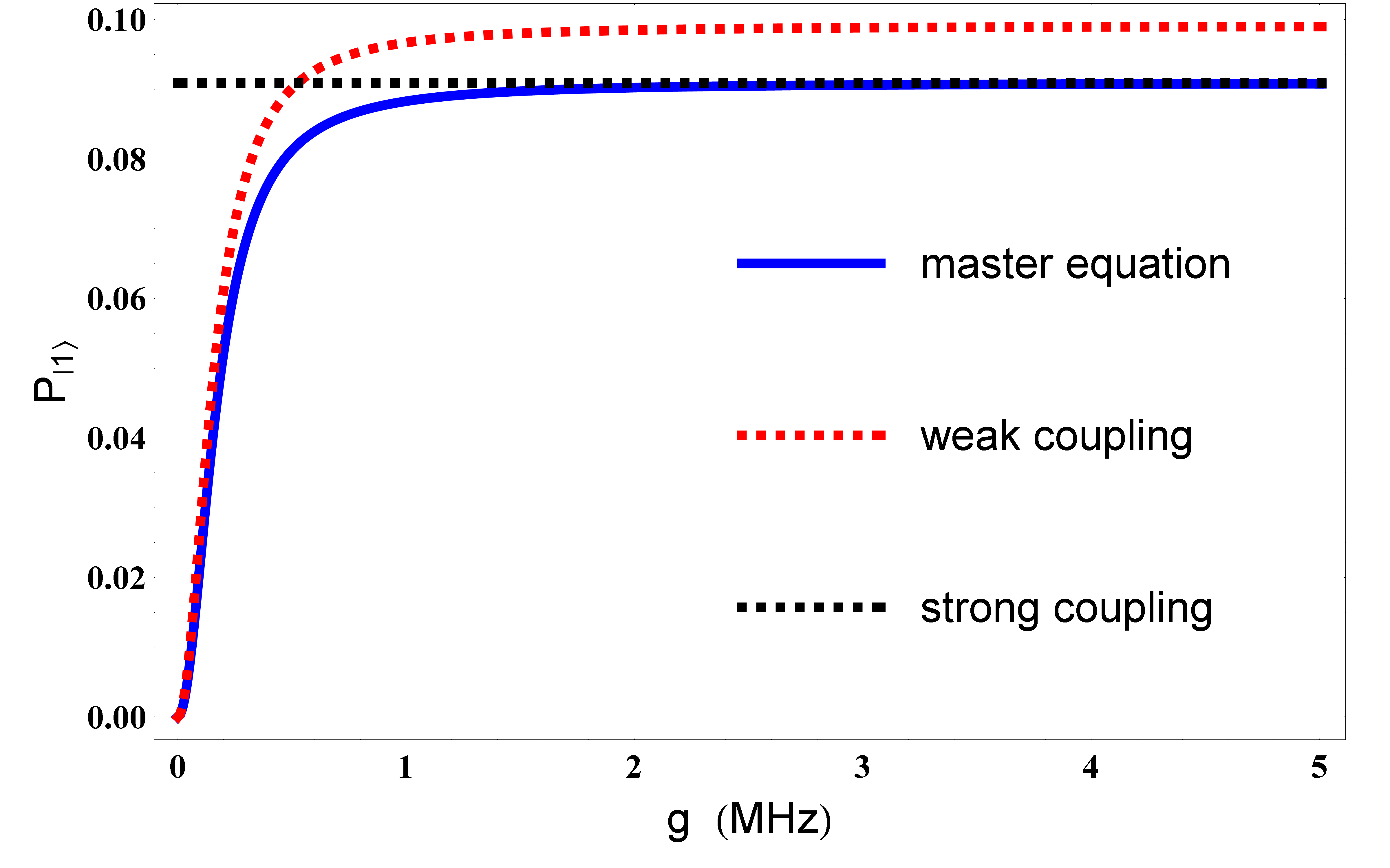}
  \label{fig:mean photon number comparison}
  }
\caption{(left) The stabilized state population and (right) mean cavity photon number for stabilization angle $\theta=\pi$, as a function of the coupling strength $g=\left< \tilde{e} 0 \right| H_{int} \left| \tilde{g} 1 \right>$, calculated with $\kappa/2\pi$= 1\,MHz, $\gamma/2\pi$ = 0.1\,MHz, $\gamma_{\phi}/2\pi$ = 0.1\,MHz. Parameters are exaggerated for enhancing visual contrast and do not reflect experimental values. At $\theta = \pi$, the rotating frame ground state overlaps with the lab frame excited state, i.e. $\left| \tilde{g} \right>=\left| e \right>$. The exact solution from the master equation coincides with the weak coupling formula when $g$ is small, whereas it falls into its asymptote predicted by the strong coupling formula when $g$ is big, showing an good agreement between the theory and the numerical calculation. \label{fig:comparison}}

\end{figure}

As for the strong coupling regime, the system first undergoes coherent oscillations between the $\left| \tilde{e} 0 \right>$ and $\left| \tilde{g} 1 \right>$ states which eventually are driven into a statistical mixture with (almost) equal population of the two levels. Therefore, at long times we can approximate the two levels as one, with a decay rate to $\left| \tilde{g} 0 \right>$ corresponding to the mean value of the cavity decay and the qubit decay $(\tilde{\gamma}_-+\kappa)/2$, as it can decay through both the qubit and the cavity channels. We can find the decay rates shown between other levels shown in Fig.~\ref{fig:weak and strong coupling regimes} in a similar fashion. Solving the corresponding optical Bloch equations again gives the stabilization fidelity as (with $\delta=\Omega_R$)

\begin{equation}
P_{\tilde{g}}=P_{\tilde{g}0}+P_{\tilde{g}1}=\frac{\tilde{\gamma}_-+\kappa}{\tilde{\gamma}_-+\tilde{\gamma}_++\kappa}.\label{eq:strong coupling fidelity}
\end{equation}

The interaction $H_{\textup{int}}$ may also induce finite coupling between other levels, through $\left< i \right| H_{\textup{int}} \left| j \right>$. However, unlike $\Gamma$ between (near-)resonant levels $\left| \tilde{e} 0 \right>$ and $\left| \tilde{g} 1 \right>$, these transition probabilities are strongly suppressed by the detuning and can be safely dropped out as long as $\left< i \right| H_{\textup{int}} \left| j \right> \ll \Omega_R,\delta$.

Eq.~\eqref{eq:weak coupling fidelity} and (\ref{eq:strong coupling fidelity}) are shown in Fig.~\ref{fig:comparison} as the two asymptotes of the fidelity versus coupling strength $g$ calculated from master equation, which shows quantitative agreements between the analytical formulas and the numerical simulations.

\section{Universal stabilization with blue-sideband interaction and Rabi drive}
In above, we have discussed how, through manipulating the Hamiltonian of a qubit-cavity system, the qubit state can be stabilized into an arbitrary superposition of its two basis. Here, as a concrete example, we demonstrate the implementation of the scheme discussed in the main text, by generating an rotating frame Hamiltonian of
\begin{equation}
H=\frac{1}{2} (\Omega_x \sigma_x+\Omega_z \sigma_z)+\Omega_b(a^\dagger \sigma^+ + a \sigma^-)+\delta a^\dagger a.\label{eq:detailed rotating frame Hamiltonian}
\end{equation}
To this end, we start from the Hamiltonian of a driven tunable coupling circuit in the lab frame dressed basis

\begin{equation}
H=\frac{\omega_q}{2}\sigma_z +\omega_c a^\dagger a+\chi a^\dagger a \sigma_z +\Omega_x \sigma_x \cos \omega_1 t+2\Omega_b(a^\dagger +a)\sigma_x \cos \omega_2 t,
\end{equation}
where the first three terms are the static energy of the device, and the last two represent the Rabi drive and the flux modulation. When flux is modulated in the vicinity of the blue-sideband frequency, this Hamiltonian can be transformed to a rotating frame by the operator $U=e^{i [\frac{\omega_1}{2}\sigma_z +(\omega_2-\omega_1) a^\dagger a] t}$ (with the fast- oscillating terms abandoned)

\begin{equation}
H_{rot}=\frac{\Omega_x}{2}\sigma_x+\frac{\omega_q-\omega_1}{2}\sigma_z+\chi a^\dagger a \sigma_z+\Omega_b(a^\dagger \sigma^+ + a \sigma^-)+(\omega_c+\omega_1-\omega_2)a^\dagger a.
\end{equation}
We immediately notice that this is equivalent to Eq.~\eqref{eq:detailed rotating frame Hamiltonian} plus an extra dispersive shift term, 

\begin{equation}
H_{rot}=\frac{1}{2}\Omega_x \sigma_x+\frac{1}{2}(\Omega_z+2\chi a^\dagger a) \sigma_z+\Omega_b(a^\dagger \sigma^+ + a \sigma^-)+\delta a^\dagger a.\label{eq:blue sideband rotating frame Hamiltonian}
\end{equation}
with $\Omega_z = \omega_q-\omega_1$ and $\delta=\omega_c+\omega_1-\omega_2$. We can approximate the above Hamiltonian as 

\begin{equation}
H_{rot}=\frac{1}{2}\Omega_x \sigma_x+\frac{1}{2}(\Omega_z+2\chi \bar{n}) \sigma_z+\Omega_b(a^\dagger \sigma^+ + a \sigma^-)+\delta a^\dagger a
\end{equation}
as long as $2\chi \bar{n} \ll \Omega_R$ is satisfied, where $\bar{n}$ is the mean cavity photon number. This requirement guarantees that the dispersive shift term can be safely counted in as only a small perturbation to the stabilization angle

\begin{equation}
\theta'=\arccos\frac{\Omega_z+2\chi \bar{n}}{\sqrt{\Omega_x^2+(\Omega_z+2\chi \bar{n})^2}}.
\end{equation}
Fig.~\ref{fig:mean photon number comparison} plots the mean cavity photon number versus the coupling strength. Similar to Fig.~\ref{fig:fidelity comparison}, in the strong coupling regime the mean cavity photon number saturates at the upper limit given by

\begin{equation}
\bar{n}_{\textup{max}}=\frac{\tilde{\gamma}_+}{\tilde{\gamma}_-+\tilde{\gamma}_++\kappa},
\end{equation}
which is small under $\kappa \gg \tilde{\gamma}_{\pm}$. In our experiment, the Rabi drive strength is 2$\pi\times$9\,MHz while the dispersive shift is less than 2$\pi\times$1\,MHz at zero dc flux, so the requirement $2\chi \bar{n} \ll \Omega_R$ is well met.

\begin{figure}
\centering
\subfloat{
  \includegraphics[width=0.45\textwidth]{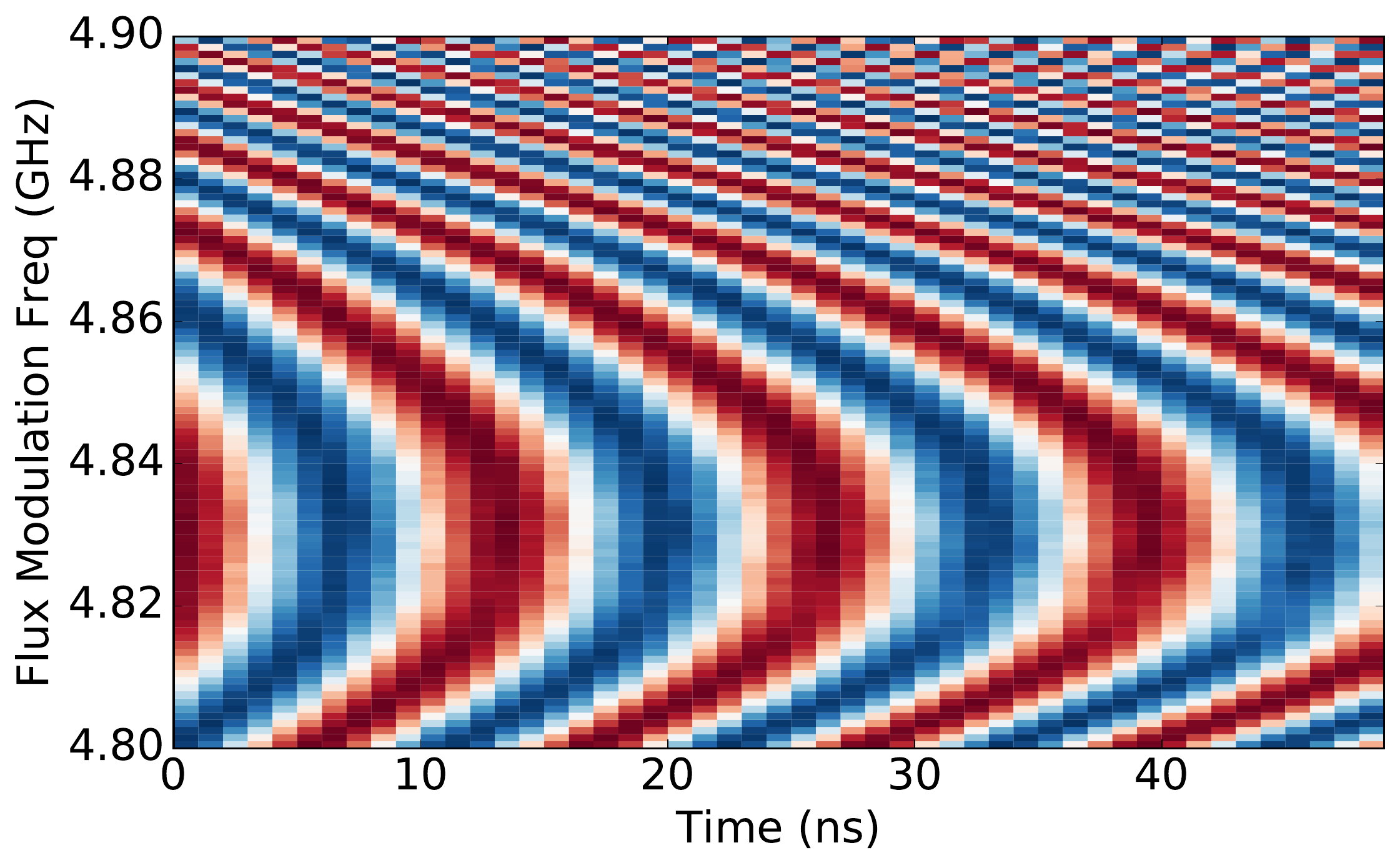}
  }
  \subfloat{
  \includegraphics[width=0.45\textwidth]{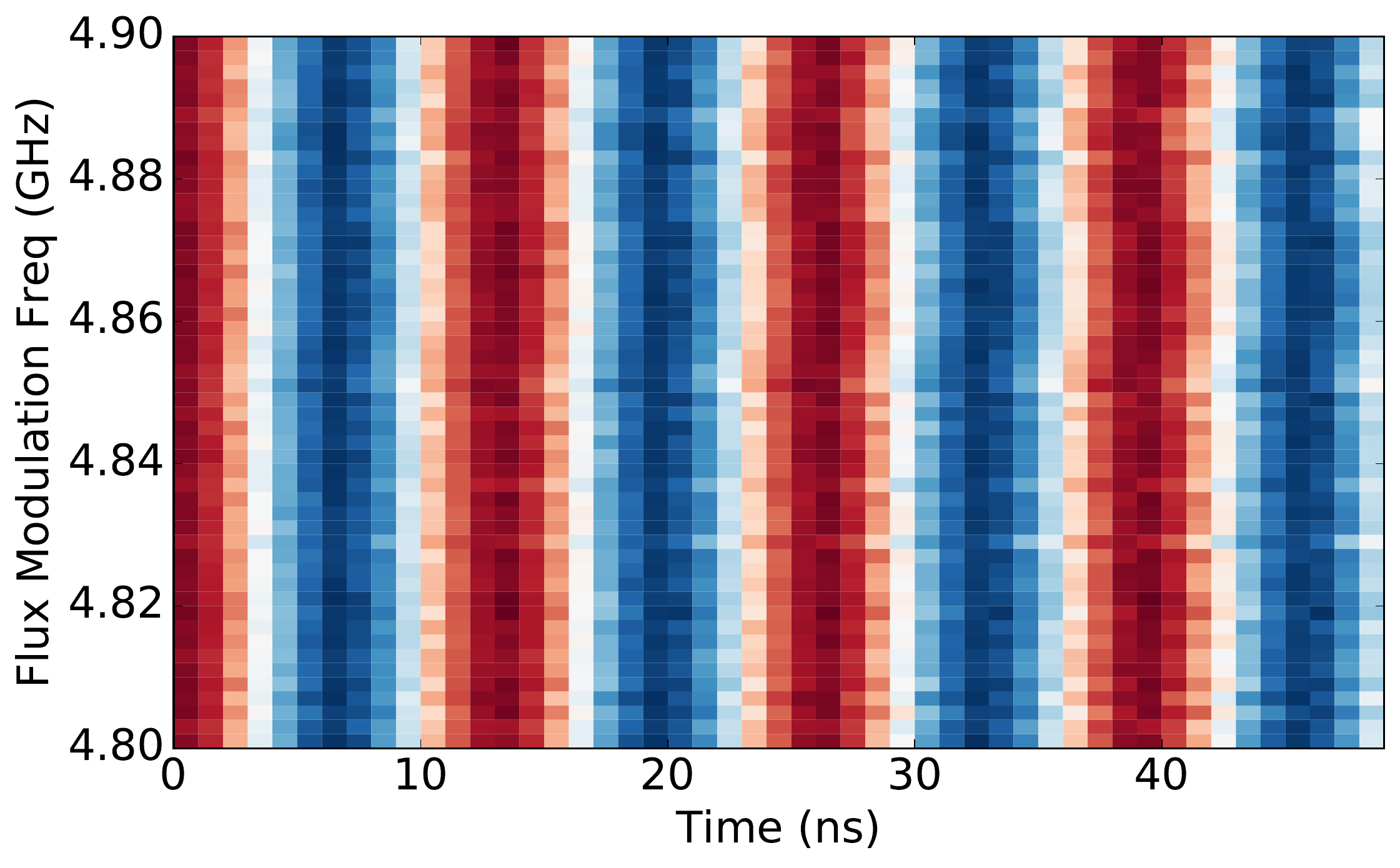}
  }
\caption{(left) Ramsey experiment done with flux modulation applied between the two qubit $\pi$/2 operations. The flux signal is created by an AWG of fixed output power. From the curved fringes qubit dc-offset/effective flux modulation strength can be measured. (right) Output power of the awg is compensated using the calibration result obtained from (left), showing vertical lines which indicates the effective flux modulation strength is kept constant over the modulation frequency range.}
\label{fig:flux calibration}
\end{figure}

\begin{figure}

\centering
\includegraphics[scale=0.6]{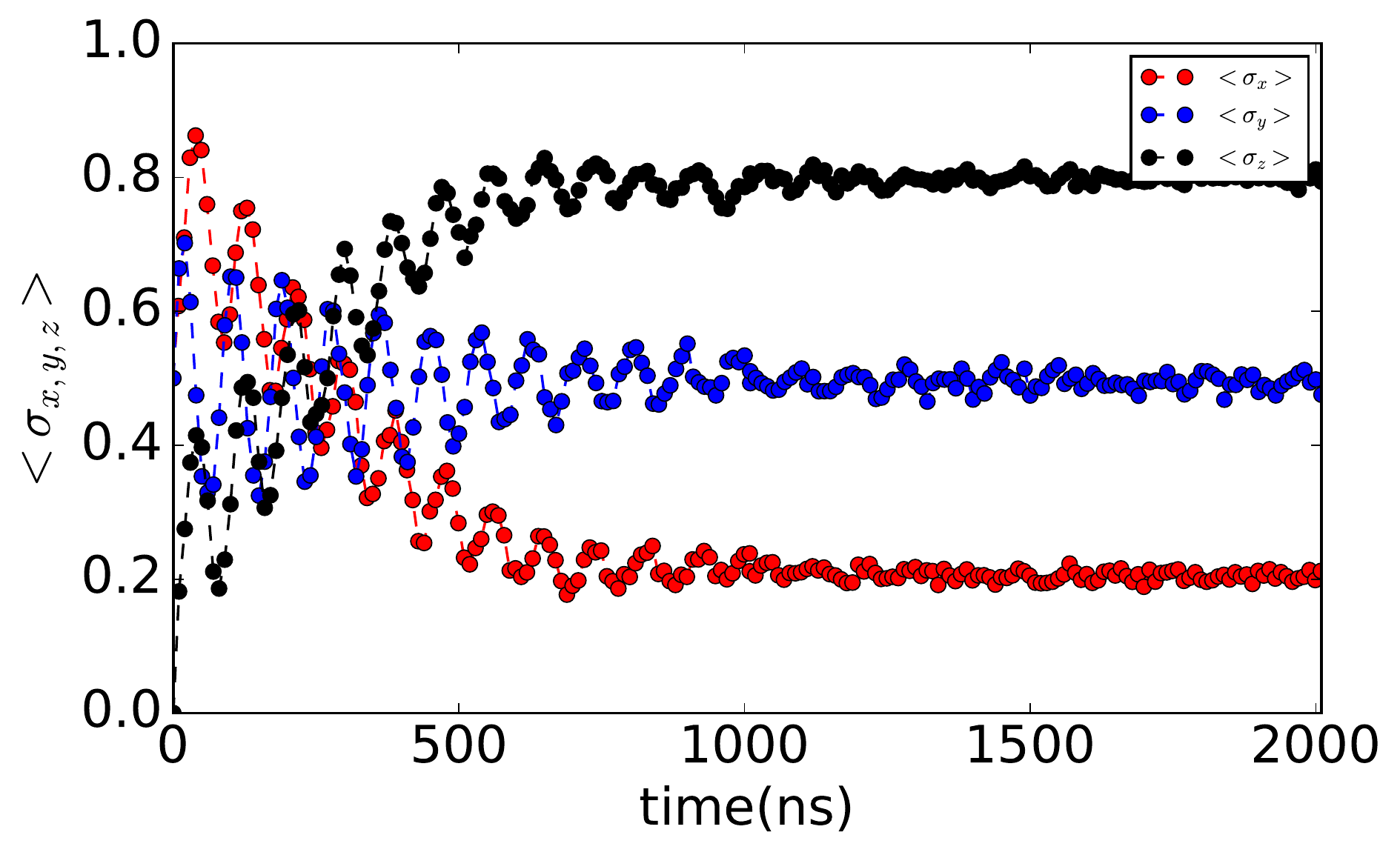}
 
\caption{Stabilization of the qubit state at polar angle $\theta = 3\pi/4$, illustrated by the evolution of its projections to the x (red), y (blue), and z (black) axis. The azimuthal angle is set to be $\phi=\pi$ to reduce overlapping between data points in this figure, by shifting the Rabi drive phase by $\pi$.}

\label{fig:xyz}

\end{figure}

The dispersive shift term can also be viewed as a qubit state dependent frequency shift to the cavity, in which sense Eq.~\eqref{eq:blue sideband rotating frame Hamiltonian} becomes
\begin{equation}
H_{rot}=\frac{1}{2}\Omega_x \sigma_x+\frac{1}{2}\Omega_z \sigma_z+\Omega_b(a^\dagger \sigma^+ + a \sigma^-)+(\delta+2\chi \sigma_z) a^\dagger a.
\end{equation}
As stated previously, the optimized fidelity is reached when $\left| \tilde{e} 0 \right>$ and $\left| \tilde{g} 1 \right>$ become degenerate, which happens at
\begin{equation}
\delta=\Omega_R-2\chi \left< \sigma_z\right> \approx \Omega_R+2\chi \cos\theta
\end{equation}
that corresponds to the qubit Rabi drive frequency $\omega_1$ and blue-sideband drive frequency $\omega_2$ being

\begin{equation}
\omega_1=\omega_q-\Omega_z, \quad \omega_2 = \omega_q+\omega_c-\Omega_R-\Omega_z.\label{eq:detunings}
\end{equation}

In our experiment, the static flux in the coupler SQUID loop is biased to zero via the dc flux line. This tunes the coupling strength $g_{r,b}$ to its minimum, allowing the blue-sideband interaction to be created by flux modulation at half of the qubit-cavity sum frequency $\Sigma$, through the second harmonic term in the Fourier series expansion of $g_b(t)$. This avoids the need to drive at the sum frequency directly, which can be technically challenging given . The blue-sideband frequency is calibrated by finding the modulation frequency that corresponds to the maximum value of the qubit excited state population, which is measured after the flux pulse is turned on for sufficiently long time. The strength of the blue-sideband interaction, $\Omega_b$, can be directly obtained from the oscillation rate of the qubit excited state population. It is a necessity for the stabilization protocol to drive the blue-sideband modulation at different frequencies depending on the stabilization target state (\eqref{eq:detunings}). However, the effective amplitude of the flux modulation will change at different frequencies, due to the frequency-dependent power loss along the rf flux line. On the other hand, the flux modulation also gives rise to a shift of the qubit frequency known as the "dc-offset", which is uniquely dependent on the modulation amplitude. As shown in Fig.~\ref{fig:flux calibration}a, Ramsey fringes can be used to directly measure the dc-offset at different flux modulation frequencies. By adjusting the output power of the arbitrary wave form generator (AWG) which is used to provide the rf flux signal, we produce a constant qubit dc-offset across the flux modulation frequency of interest (Fig.~\ref{fig:flux calibration}b), equivalent to realizing a constant blue-sideband interaction strength for all these frequencies. The strength of the Rabi drive $\Omega_x$, which is also kept fixed throughout the stabilization protocol, can be directly measured from the Rabi experiment. 

The Rabi drive pulse and the blue-sideband flux pulse are simultaneously sent to the circuit sample, with detunings $\Omega_z$ and $\Omega_z+\Omega_R$, respectively. Qubit tomography, with phase synchronized to the Rabi drive, is performed at different pulse times. As is displayed in Fig.~\ref{fig:xyz}, a coherent oscillation of the qubit state is observed at the beginning of time, with a rate close to the total Rabi rate $\Omega_R$. We set the initial phase of the Rabi drive to zero (for Fig.~\ref{fig:xyz} it is set to $\pi$), so that in the long-time limit the qubit state will be stabilized with $\left<\sigma_y\right> \approx 0$, while the polar angle and the purity are measured as 

\begin{equation}
\theta_{\rm measured}=\arccos\frac{\left<\sigma_z\right>}{\sqrt{\left<\sigma_x\right>^2+\left<\sigma_y\right>^2+\left<\sigma_z\right>^2}},
\end{equation}

\begin{equation}
\left|\left<\vec{\sigma}\right>\right| = \sqrt{\left<\sigma_x\right>^2+\left<\sigma_y\right>^2+\left<\sigma_z\right>^2}.
\end{equation} 

\section{A more efficient scheme with the ``purple" sideband}

\begin{figure}
\centering
\subfloat{
  \includegraphics[width=0.32\textwidth]{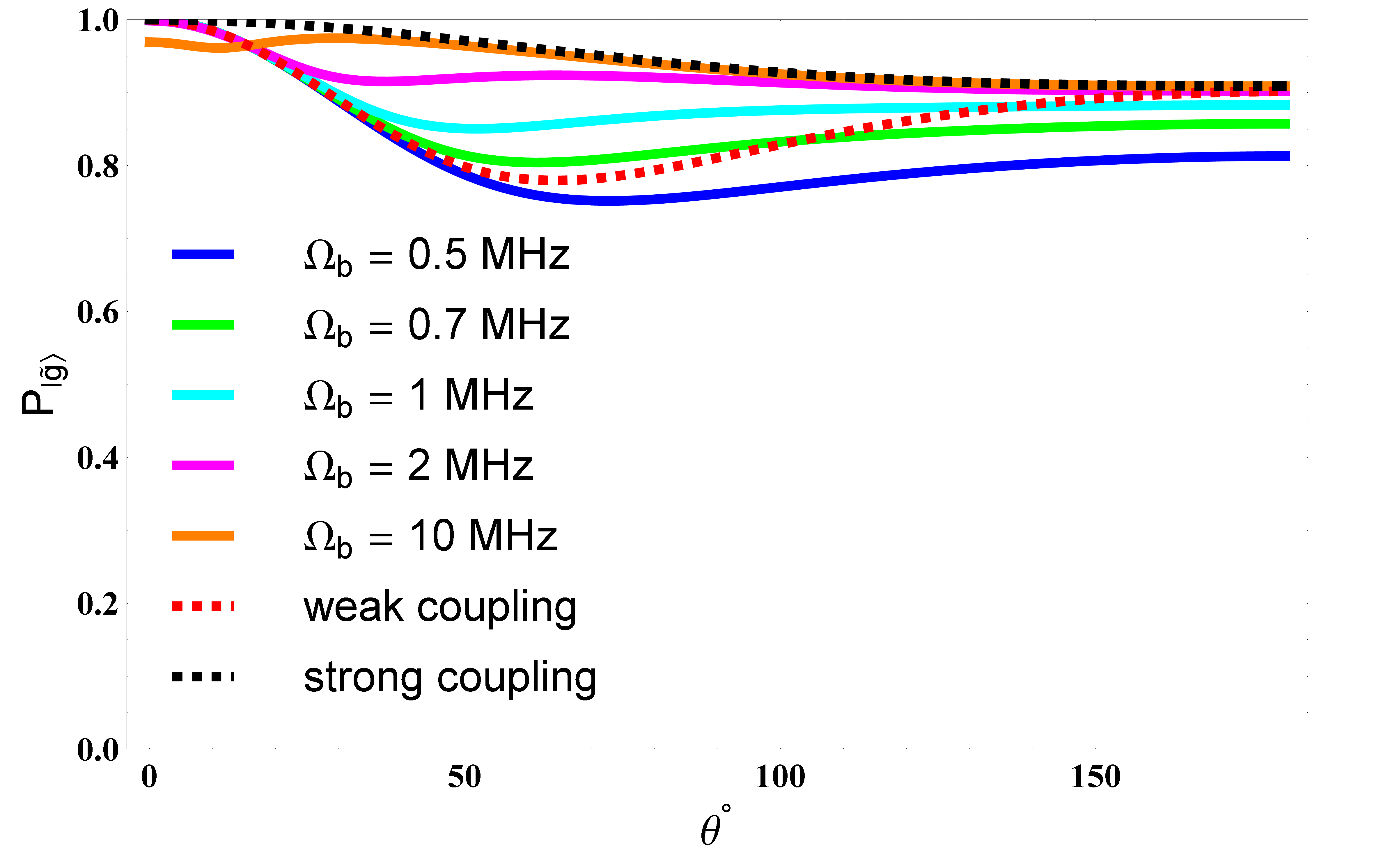}
  \label{fig:blue universal}
  }\hfill
  \subfloat{
  \includegraphics[width=0.32\textwidth]{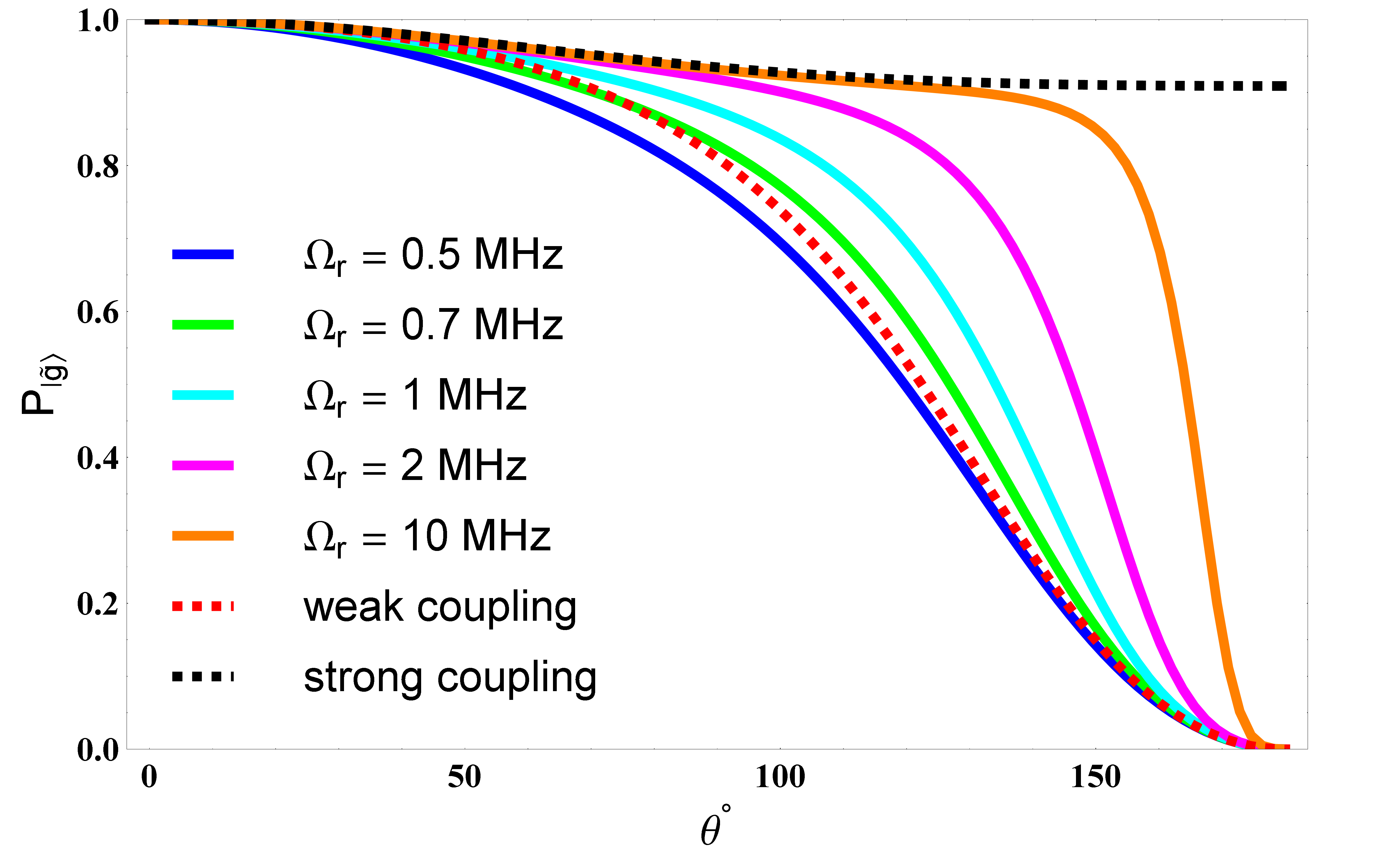}
  \label{fig:red universal}
  }\hfill
    \subfloat{
  \includegraphics[width=0.32\textwidth]{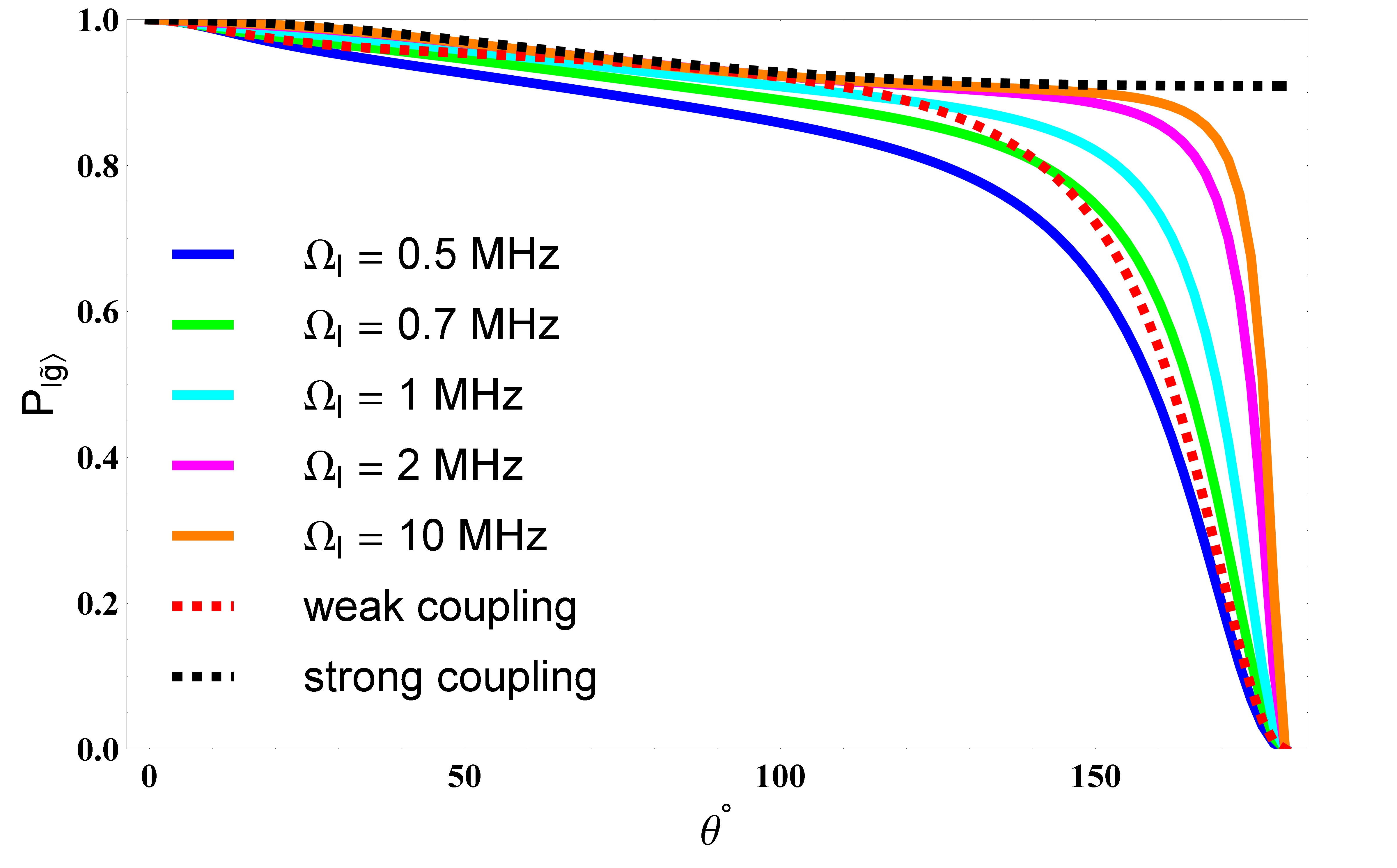}
  \label{fig:Longitudinal universal}
  }
 
\caption{Comparison of stabilization fidelity with respect to  $P_{\tilde{g}}$ at different polar angles $\theta$, between schemes using (left) blue-sideband interaction $\Omega_b(a^\dagger \sigma^+ + a\sigma^-)$, (center) red-sideband interaction $\Omega_r(a^\dagger \sigma^- + a\sigma^+)$ and (right) longitudinal interaction $\Omega_l(a^\dagger+a)\sigma_z$. All three interactions can be realized with the tunable coupling circuit, by using flux modulations and charge drives. Each scheme is calculated with different values of their coupling strength, ranging from the weak coupling regime to the strong coupling regime. The red and black dashed lines are theoretical limitations in the weak ($\Omega_{b,r,l}/2\pi = 0.5\,$MHz) and the stronger coupling regime ($\Omega_{b,r,l}/2\pi = 10\,$MHz), respectively. For all three schemes, the increase of the interaction strength results in higher overall fidelity levels, gradually approaching the upper limit set by Eq.~\eqref{eq:strong coupling fidelity}. However only the blue-sideband interaction allows for universal stabilization throughout all $\theta$ values, as it uniquely remains highly efficient up to $\theta=\pi$ when the other two rapidly lose fidelity.}

\label{fig:interaction comparison}

\end{figure}

There are several types of interactions that can all be conveniently realized by the tunable coupling circuit. For example, flux modulation at the red- or the blue-sideband frequency will result in their correspond sideband interactions, and Rabi drive through qubit's charge port at the cavity's frequency will lead the longitudinal interaction. The reason why we choose the blue-sideband interaction  for our scheme is well explained by Fig.~\ref{fig:interaction comparison}, which shows the comparison of the stabilization performance between schemes using different types of interactions. While all three schemes could attain stabilization with high efficiency at small $\theta$, only the blue-sideband interaction is able to couple $\ket{g0}$ and $\ket{e1}$, which is critical for preserving a good fidelity up to $\theta=\pi$.

\begin{figure}

\centering
\includegraphics[scale=0.6]{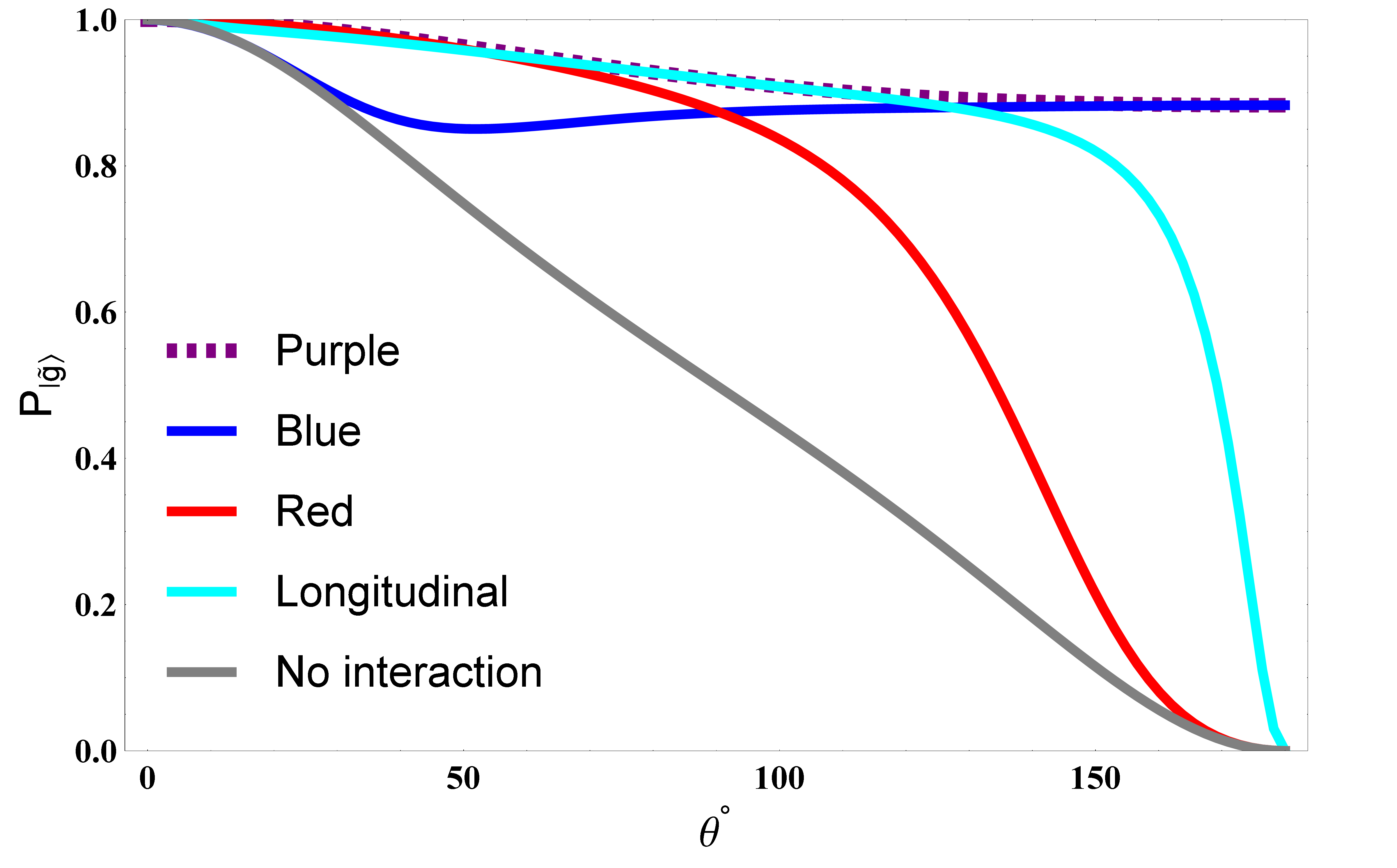}
 
\caption{A comparison of stabilization schemes with different interactions, by plotting their stabilization state population $P_{\tilde{g}}$ as a function of the stabilization angle $\theta$. The grey curve represents ``stabilization" from qubit's natural decay without interactions at play. All interaction terms have the same coupling strength of $2\pi\times 1$\,MHz. Other parameters are $\Omega_R/2\pi=100$\,MHz, $\kappa/2\pi=$1\,MHz, $\gamma/2\pi=$0.1\,MHz and $\gamma_{\phi}/2\pi=$0.1\,MHz. The purple interaction outperforms all of the other interactions by providing highest stabilization population fidelities for all angles.}

\label{fig:optimal interaction}

\end{figure}

An intuitive impression can be gained from Fig.~\ref{fig:interaction comparison} as well, that by mixing multiple interactions together, the weakness of one interaction can be compensated by the other, which promises a truly universal scheme with higher fidelities. The optimal interaction term that is universally efficient for all stabilization angles  is simply given by
\begin{equation}
H_{\rm int}=\Omega_P(a^\dagger+a)(e^{i\phi}\tilde{\sigma}^++e^{-i\phi}\tilde{\sigma}^-)
\end{equation}
where $\phi$ is an arbitrary phase, and $\tilde{\sigma}^+$ is defined as
\begin{equation}
\tilde{\sigma}^+=\left|\tilde{e}\right>\left<\tilde{g}\right|.
\end{equation}
Transforming it back to the lab basis through the unitary operator from Eq.~\eqref{eq:unitary transformation},
\begin{equation}
\tilde{\sigma}^+=\frac{1}{2}\begin{pmatrix}
\sin\theta&-1+\cos\theta\\ 
1+\cos\theta&-\sin\theta
\end{pmatrix},
\end{equation}
so
\begin{equation}
e^{i\phi}\tilde{\sigma}^++e^{-i\phi}\tilde{\sigma}^-=\begin{pmatrix}
\cos\phi\sin\theta&-i\sin\phi+\cos\phi\cos\theta\\ 
i\sin\phi+\cos\phi\cos\theta&-\cos\phi\sin\theta
\end{pmatrix}.
\end{equation}
Something truly magical will happen, a mathematical accident or a deep and beautiful
piece of physics depending on how one looks at it, if we set $\phi=\pi/2$ here: we then simply arrive to the ``purple'' sideband interaction
\begin{equation}
H_{\rm int}=\Omega_P(a^\dagger+a)\sigma_y
\end{equation}
which is a balanced mixture of the red- and the blue-sideband interactions completely independent of $\theta$. Fig.~\ref{fig:optimal interaction} displays the comparison between the purple-sideband stabilization and the other three schemes, which shows that under the same coupling strength, the purple-sideband interaction provides the highest stabilization fidelity at all angles. This interaction can be generated by driving the tunable coupling device  at the red- and blue-sideband frequency simultaneously with equal drive strength.

\end{document}